\documentclass{IEEEtran}

\usepackage{graphicx}
\usepackage{amsmath}
\usepackage{amssymb}
\usepackage[mathscr]{eucal}
\usepackage{bm}
\usepackage{amscd}

\newcommand{\beq}{\begin{equation}}
\newcommand{\eeq}{\end{equation}}
\newcommand{\beqa}{\begin{eqnarray}}
\newcommand{\eeqa}{\end{eqnarray}}
\newcommand{\beqan}{\begin{eqnarray*}}
\newcommand{\eeqan}{\end{eqnarray*}}

\newcommand{\tr}[1]{{\rm tr} \left( #1 \right) }
\newcommand{\ket}[1]{| #1 \rangle}
\newcommand{\bra}[1]{\langle #1 |}
\newcommand{\braket}[2]{\langle #1 | #2 \rangle}

\newcommand{\llangle}{\langle\!\langle} 
\newcommand{\rrangle}{\rangle\!\rangle}

\newcommand{\ad}{{\rm ad}}

\newcommand{\Hi}{\mathcal{H}}
\newcommand{\Hin}{\mathcal{H}^N}
\newcommand{\Si}{\mathcal{S}}
\newcommand{\Ei}{\mathcal{E}}

\newcommand{\supp}{\textrm{supp}}

\newcommand{\herm}{\mathfrak{H}}
\newcommand{\bound}{\mathfrak{B}}

\newcommand{\eneig}{e}

\newcommand{\C}{\mathbb{C}}
\newcommand{\N}{\mathbb{N}}
\newcommand{\R}{\mathbb{R}}

\newcommand{\E}{\mathbb{E}}

\newcommand{\Li}{\mathcal{L}}
\newcommand{\LL}{\bm{L}}
\newcommand{\ddt}{\frac{d}{d t}}

\newcommand{\diag}{\textrm{diag}}

\newcommand{\trace}{\textrm{tr}}
\newcommand{\um}{\frac{1}{2}}

\def\supp{{\rm supp}\ }

\newtheorem{theorem}{Theorem}

\newtheorem{proposition}{Proposition}
\newtheorem{lemma}{Lemma}
\newtheorem{corollary}{Corollary}
\newtheorem{remark}{Remark}
\newtheorem{example}{Example}

\begin{document}


\title{Modeling and Control of Quantum Systems: An Introduction}

\author{Claudio Altafini\thanks{
C. Altafini is with SISSA,
International School for Advanced Studies,
via Bonomea 265, 34136 Trieste, Italy
(\texttt{altafini@sissa.it})} and
Francesco Ticozzi \thanks{F. Ticozzi is with the Dip. di Ingegneria dell'Informazione, Universit\`a di Padova, via Gradenigo 6/B, 35131 Padova, Italy ({\tt ticozzi@dei.unipd.it}) and Dept. of Physics and Astronomy, Dartmouth College, Wilder 6127, 03755 Hanover, NH (USA)}}

\maketitle 

\begin{abstract} 
The scope of this work is to provide a self-contained introduction to a selection of basic theoretical aspects in the modeling and control of quantum mechanical systems, as well as a brief survey on the main approaches to control synthesis. While part of the existing theory, especially in the open-loop setting, stems directly from classical control theory (most notably geometric control and optimal control),
a number of tools specifically tailored for quantum systems have been developed since the 1980s, in order to  take into account their distinctive features: the probabilistic nature of atomic-scale physical systems, the effect of dissipation and the irreversible character of the measurements have all proved to be critical  in feedback-design problems. The relevant dynamical models for both closed and open quantum systems are presented, along with the main results on their controllability and stability. A brief review of several currently available control design methods is meant to provide the interested reader with a roadmap for further studies.

\end{abstract}



\section{Introduction}\label{introduction}


Quantum dynamical systems model the evolution characterizing physical phenomena at atomic scales. 
As the effective capabilities of manipulating matter and light at those scales is steadily growing and new experimental and technological possibilities are at reach, the interest in developing systematic theories for {\em controlled quantum mechanical systems} is rapidly increasing. In addition, the potential advantage offered by {\em quantum information processing} \cite{nielsen-chuang} has given to quantum control a new perspective, and a new set of tasks: the ability of employing quantum systems to store, manipulate and retrieve information requires an unprecedented degree of control, and further motivates the development of control schemes specifically tailored to the quantum mechanical setting.
Since the time of the first pioneering contributions to the field \cite{bs12,Huang1983,Belavkin1983,Peirce1988}, quantum control methods have gained a stable role in a number of experimental settings, among which we recall:

\noindent{\em Laser-driven molecular reactions} -- Thanks to control-theoretic techniques (in particular optimal control),  the design of laser fields able to induce selected molecular reactions has evolved from an heuristic to a systematic approach \cite{Tannor2,Peirce1988,brumer92laser,Brif2010Control}.
Optimal control in the context of reaction dynamics and other aspects of the problem are reviewed {\em e.g.} in \cite{Peirce1988,Dahleh1,Rabitz1}.

\noindent{\em Pulse sequences design in Nuclear Magnetic Resonance (NMR)} -- In NMR spectroscopy, the system (an ensemble of atomic or nuclear spins) is typically steered by means of sequences of strong electromagnetic pulses \cite{Ernst1,Mehring1,Cory1modif}. The bewildering range of open loop control methods developed in NMR over the last 60 years has, until only recently \cite{viola-seminalDD,Khaneja3}, passed unnoticed by the control community. Alongside piecewise constant controls, these methods include also averaging techniques and ``chattering'' control \cite{Haeberlen1}.

\noindent{\em Adaptive quantum measurements} -- the first type of closed-loop quantum control to be considered theoretically was motivated by the engineering of optimal detector for classical communication over quantum media \cite{dolinar,helstrom}.
Following those ideas, theoretically advanced adaptive measurement experiments also based on quantum feedback control experiments have been performed \cite{wiseman-adaptive,armen,geremia},
and this continues to be a very active research area.

\noindent{\em Feedback control of optical systems} -- Quantum-optical systems and networks represent a natural platform to implement feedback control methods: a significant body of theoretical and experimental work has been developed since the early experimental and theoretical contributions on control of traveling-wave quantum fields \cite{wj85,hy86,yim86,ssh}, which were among the first quantum-limited feedback control experiments to be performed. A number of interesting, control oriented results have been concerned with control of simple systems trapped in optical cavities by designing a laser beam, with a focus on state preparation and cooling problems \cite{qed1, qed2,orozco1,vanhandel-feedback}. We refer the reader to the monograph \cite{wiseman-book} for a comprehensive review, including historical perspectives, from the quantum optics viewpoint.

Other systems for which control-theoretic methods are being successfully employed include {\em quantum dots} \cite{petta-science-mod,qdots1}, {\em opto- and nano-mechanical resonators} \cite{nanomechanical1,nanomechanical2,nanomechanical3}, {\em trapped ions} \cite{ions1,ions2-mod}, {\em Josephson-junctions} \cite{josephson1, josephson2}, {\em cold atoms} \cite{cold0,cold1} and, recently, {\em nitrogen-vacancy centers in diamonds} \cite{Jacques09-mod,Jiang09-mod}. Discrete-time approaches to quantum measurements and feedback control have been also used for experiments on single photons \cite{gillet-mod} and spin ensembles \cite{polzik,hammerer}.

\noindent Such a diverse range of applications makes it quite clear that an exhaustive introduction to the field of quantum control cannot arguably be condensed in the form of a tutorial paper. Engaged in this endeavor, we have inevitably been forced to face some difficult choices. We opted for a selection of topics that can be presented to a {\em control engineering} audience avoiding long technical detours, and that in part complement other introductory material available {\em e.g.} in  \cite{Dahleh1,Mabuchi2005,dalessandro-book, wiseman-book,vanhandel-invitation,Dong-Petersen-survey,Brif2010Control}.
We chose not to describe any particular, physically-relevant example in detail, since each experimental system presents a number of particular features that cannot be thoroughly discussed in the limits of this work. In the spirit of systems and control theories, we would like instead to provide the reader with a set of theoretical tools that are as ``portable'' (from an experimental setting to another) as possible. This of course comes with a cost: in order to move to applications, one has to carefully verify the applicability of the results, and most likely adapt them to the peculiarities of his/her own setting. The models we discuss are always {\em continuous in time}: this of course leaves out a significant number of interesting results, but allows for a more consistent presentation and, in our opinion, a closer connection to classical control theory. 

Some of the topics related to measurement and feedback could certainly be presented in a quantum probability framework in a more rigorous way, as it has been recently done by part of the quantum control community: nonetheless, this approach may appear intimidating to a newcomer to the field, rendering the connection with earlier results less direct, and it is not the most natural for investigating {\em e.g.} controllability issues. We therefore introduce and use notions directly stemming from it (as {\em e.g.} quantum conditional expectations) only when needed, and in a somewhat simplified manner, referring to the extensive literature for thorough expositions of these ideas. The emerging connections with (classical) stochastic models and control are highlighted when appropriate. 

The paper is structured as follows: The first part (Section \ref{sec:essentials}--\ref{sec:open}) is intended as a concise tutorial on quantum controlled dynamical systems. More in detail, Section \ref{sec:essentials} presents the basic (static) tools for building a quantum system (states, observables and ways to compute probabilities), Section \ref{sec:dynamics} introduces the dynamical models for isolated quantum systems and Hamiltonian control. In Section \ref{sec:open} the reader is introduced to the three main approaches to the description of controlled open systems (in continuous time): Hamiltonian description, master equations and stochastic models. 

The second part is devoted to control-oriented issues. Section \ref{sec:controllability} guides the reader through the  fundamental controllability analysis of the deterministic models. The rest of the paper (Section \ref{sec:openloop}-\ref{sec:closedloop}) should serve as a brief survey of the wide spectrum of synthesis methods that have been presented in the literature, including both open- and closed-loop methods. The authors' perspective on the next challenges in the field is subsumed in Section \ref{outlook}.


\section{Quantum essentials:\\ Observables, States and Probabilities}\label{sec:essentials}

This section is devoted to present the basic elements of quantum theory for {\em finite-dimensional systems}. For more details, and other modern approaches to quantum theory, we refer the reader to the monographs \cite{sakurai,Peres2,isham-qm}. To every quantum system ${\cal Q}$ is associated a complex Hilbert space $\Hi.$ The dimension of the Hilbert space depends on the variables of the system we aim to describe: here, for the sake of simplicity and in order to avoid technical complications, we will only consider finite-dimensional quantum systems $\Hi=\Hi^N\simeq \C^N$, namely systems whose variable of interest can assume only a finite number of outcomes. Elementary examples are quantum models describing spin or polarization properties \cite{sakurai}. Finite dimensional systems also allow for simple linear-algebraic representations, helping to establish explicit connections with classical system-theoretic methods, and are of key interest in the emerging field of quantum information \cite{nielsen-chuang}.

\subsection{Observable quantities}

Any (real-valued) physical variable for the system, or {\em observable} in the physics terminology, is associated to an Hermitian operator $Y\in\herm(\Hin).$  By the spectral theorem, $Y$ admits a representation $Y=\sum_jy_j\Pi_j,$ where $\{y_j\}\subset \R$ are the eigenvalues of $Y$ and the corresponding orthogonal projectors $\{\Pi_j\}$ form a resolution of the identity, namely $\Pi_k\Pi_j=\delta_{kj}\Pi_j,$ $\sum_j\Pi_j=I$. 
The eigenvalues $\{y_j\}$ then represent the {\em possible outcomes} of a measurement of $Y$, and the $\Pi_j,$ which play the role of {\em quantum events}, let us compute the corresponding probabilities, given the state of the system. 

To learn more about the non-commutative probability framework that encompasses quantum mechanics, see Remark 1. For a thorough exposition, we refer to \cite{parthasarathy}, or to the excellent introduction \cite{maassen-qp}.

\subsection{State vectors}
When the state of the system is (ideally) known exactly, it can be described by a {\em state vector} $ \ket{\psi} $ (see Appendix \ref{notations} for a review of Dirac's notation). $ \ket{\psi}$ is a norm-1 vector in the complex $N$-dimensional Hilbert space $ {\cal H}$. 

A particular choice of basis is employed throughout this tutorial: consider the observable $ H_0 = \sum_j \eneig_j \Pi_j \in \herm(\Hi) $ associated to the {\em energy of the system}, that is, the {\em Hamiltonian} of the system, and choose an orthonormal basis of eigenvectors of $H_0$, $\{\ket{\eneig_1},\ldots,\ket{\eneig_N}\}.$ In this basis, we can represent $H_0$ as a diagonal matrix:
\beq
H_0 = {\rm diag} (\eneig_1, \ldots,  \eneig_N ).
\label{eq:H0}
\eeq
If we write the state vector of the system with respect to this basis,
$\ket{\psi} = c_1 \ket{\eneig_1} + \ldots + c_N \ket{\eneig_N}$, then 
\beq
\braket{\psi}{\psi} =  \sum_{j=1}^N | c_j | ^2 =1,
\label{eq:sum=1}
\eeq 
and $ | c_j | ^2 $ is called the population of the $j$-th eigenstate.  
Geometrically, \eqref{eq:sum=1} means that the state vector $ \ket{\psi} $ is living on the unit sphere on the Hilbert space ${\cal H} $ of dimension $N$: $ \ket{\psi} \in \mathbb{S}^{2N-1} \subset {\cal H} $.

\subsection{Density operators}

Density operators are used to describe the state of {\em statistical ensembles} i.e., collections of identical quantum systems, or of a single system in the presence of classical uncertainty. More precisely, assume that $f_j,$ $j=1,\ldots,m$, is the fraction of population of some ensemble prepared in the state $ \ket{\psi_j},$ with different state vectors not necessarily orthogonal to each other. The associated quantum {\em density operator} is defined by 
\beq
\rho = \sum_{j=1}^m f_j  \ket{\psi_j}\bra{\psi_j } \qquad \text{ s.t. } \quad f_j \geq 0 , \quad \sum_{j=1}^m f_j =1. 
\label{eq:def-dens2}
\eeq 

While to every ensemble $\{f_j,\ket{\psi_j}\}$ corresponds a unique $\rho,$ the same density operator can emerge from different ensembles: for example, $\rho=\frac{1}{N} I$ can be obtained by any ensemble with $f_j=\frac{1}{N},\,\forall j$ and $\{\ket{\psi_j}\}$ any orthonormal basis. A density operator can also be used to describe the state of a single system prepared in an uncertain state, much alike a probability density in the classical (Bayesian) framework. In general, the density operator is a square complex matrix of dimension $N\times N$ such that 
\[1)\; \rho= \rho^\dagger \geq 0 ;
\quad 2)\;  \tr{\rho}= 1 ;
\quad 3)\; \tr{\rho^2}\leq 1. \]
Properties 1) and 2) are straightforward given \eqref{eq:def-dens2} and invoking the linearity of the trace, while 3) is directly implied by 1) and 2). The quantity $ \tr{\rho^2 } $ is called the {\em purity} of $ \rho$. Call $ \mathfrak{D}(\Hi) $ the set of matrices satisfying 1)-2) above:
\[ \mathfrak{D}(\Hi)=\{\rho\in{\mathfrak{B}}(\Hi)\,|\,\rho=\rho^\dag\geq 0,\,\tr{\rho}=1\}.\]
The structure of $\mathfrak{D}(\Hi) $ is not easy to describe: the constraint $ \tr{\rho}=1 $ makes the convex cone $ \rho= \rho^\dagger \geq 0$ into a compact and convex set. 
We will see below that for $N=2 $ the structure of $\mathfrak{D}(\Hi)$ can be made explicit by means of the so-called Bloch vector representation. An important subset of density operators is the following: if $f_1=1,$  the whole ensemble is prepared in the same state $ \ket{\psi} ,$ so that $ \rho = \ket{\psi}\bra{\psi}$ is a rank-one orthogonal projector. Such a $\rho$ is called a {\em pure state}. 
In this case, the description one obtains is completely equivalent to that provided by the state vector $ \ket{\psi},$ up to an irrelevant global phase (see below). 
On the other hand, an ensemble in which at least two of the $f_j $ of eq. \eqref{eq:def-dens2} are nonzero is called a mixed ensemble, or {\em mixed state} and does not admit a description in terms of a single state vector. 
In matrix representation, it is always possible to apply a change of basis that diagonalizes $ \rho $. Then we have
$
\rho = {\rm diag} ( \mu_1 , \ldots, \mu_N ) $, 
where $ \mu_1, \ldots \mu_N $ are the eigenvalues of $ \rho$, $ \mu_j \geq 0$, $j=1, \ldots N$, $\sum_{j=1}^N \mu_j =1$.
In particular, we have that, up to a permutation of the eigenvalues,
for {\em pure states}: $ \mu_1 =1 , \; \mu_2=\ldots = \mu_N =0 $ and hence $ \tr{\rho^2}=1$;
while for generic {\em mixed states} at least two eigenvalues are nonzero $ \mu_1 \neq 0 $, $ \mu_2 \neq 0$ and $ \tr{\rho^2 } < 1$.
The state with $ \mu_1 =\mu_2 = \ldots = \mu_N = \frac{1}{N} $ is called the {\em completely mixed state} (or completely random state). It is characterized by $ \tr{\rho^2} = \frac{1}{N} $.

\subsection{Probabilities and expectations}

If the state of a system is described by $\ket{\psi},$ the probability of obtaining $y_j$ as an outcome of $Y=\sum_jy_j\Pi_j$ is computed as
\beq
\label{wf:probability} p_j=\bra{\psi}\Pi_j\ket{\psi}.
\eeq
Thus $p_j$ is real and positive, and such that
$ 
\sum_j p_j
=\bra{\psi}\sum_j\Pi_j\ket{\psi} 
=\braket{\psi}{\psi} 
=1$. 
The state vector, after a measurement outcome has been recorded, becomes 
\beq\label{wf:conditioning}\ket{\psi|_{Y = y_j}}=\frac{\Pi_j\ket{\psi}}{\bra{\psi}\Pi_j\ket{\psi}^\um}, \eeq 
where $\ket{\psi|_{Y = y_j}}$ denotes the state vector conditioned on recording the outcome $y_j$ in a measurement of the observable $Y$.
It is worth noticing that the global phase of the state vector does not carry physical information, that is $\ket{\psi}$ and $e^{i\phi}\ket{\psi}$ lead to the same predictions, according to \eqref{wf:probability}--\eqref{wf:conditioning}, for any $\phi\in\R$.


For density operators, the prediction and conditioning rules \eqref{wf:probability}--\eqref{wf:conditioning}, along with the fact that $\bra{\psi}X\ket{\psi}=\trace(\ket{\psi}\bra{\psi}X)$ for any $X\in\herm(\Hi^N),$ imply that for a pure state, given an observable $Y=\sum_jy_j\Pi_j,$ the probability of obtaining $y_j$ as an outcome of $Y$ is computed as
\beq\label{d:probability} p_j=\trace(\rho\Pi_j);\eeq
and the conditional density operator after the outcome $y_i$ has been recorded becomes 

\beq\label{d:conditioning}\rho|_{Y=y_j}=\frac{\Pi_j\rho\Pi_j}{\trace(\rho\Pi_j)}. \eeq 
On mixed states \eqref{d:probability} still holds by linearity, and the validity of \eqref{d:conditioning} is extended to this generic case.

Assuming that the state before the measurement is $\rho$, we want to compute the expectation value of an observable $Y$. Starting by the (classical) definition of expectation and using the linearity of the trace, we get that
\[\E_{\rho}(Y):=\sum_jy_jp_j=\sum_jy_j\trace(\Pi_j\rho)=\trace(Y\rho).\]
Hence, the expectation value of observables can be computed directly as the Hilbert-Schmidt inner product of $Y$ and $\rho.$ 

Let us remark that the conditioning rule $\eqref{d:conditioning}$ directly leads to a remarkable difference between classical and quantum probability. In fact, suppose a measurement of $Y$ has been performed, but no information on the outcome is available (a situation sometimes referred as non-selective measurement). Thus one can compute the ``average'' state:
\[\bar\rho=\sum_jp_j\rho|_{Y=y_j}=\sum_j\Pi_j\rho\Pi_j.\]
This is in general different from $\rho$, in striking contrast with the classical case.

\begin{remark}[Retrieving classical probability]\label{classprob}
Assume we fix an observable of interest with non-degenerate spectrum, say $Y=\sum_{j=1}^N y_j\Pi_j,$ and consider only the other operators that commute with $Y,$  $[X,Y]=[\rho,Y]=0.$ Thus in spectral form we have $\rho=\sum_j\mu_j\Pi_j,$ $X=\sum_jx_j\Pi_j.$ Any valid event is associated to an orthogonal projection $\Pi$ that can be written as $\Pi=\sum_j\pi_j\Pi_j,$ with $\pi_j\in\{0,1\}.$
Then all these operators are completely specified by their spectra, which are functions from $\Omega=\{1,\ldots,N\}$ to $\R$. Let us now interpret $\Omega$ as an abstract sample space, and define the probability distribution
${\mathbb P}(\omega)= \mu_\omega$ associated to the spectrum of $\rho$.
%
The ``classical'' rules are then retrieved from the quantum ones: the probability of an event $\Pi$ can be computed through its spectrum $\pi$, playing the role of the indicator function: 
\[\tr{\rho \Pi}=\sum_{j=1}^N\tr{\Pi_j\Pi}\mu_j=\sum_{\omega\in\Omega} \pi_\omega{\mathbb P}(\omega)={\mathbb P}(\Pi).\]
Similarly, we can compute expectation values of observables interpreting their spectra as random variables:
\[\tr{\rho X}=\sum_jx_j\mu_j=\sum_\omega x_\omega{\mathbb P}(\omega)=\E(X).\]
Building on these ideas, one can rigorously ``encode'' a classical probability space in a commutative algebra, and then obtain a quantum probability space by removing the commutativity constraint \cite{maassen-qp}. From this viewpoint, the quantum two-level system that we will detail in the next section can be thought as the non-commutative version of a classical coin.
\end{remark}

\subsection{Two-level systems and Bloch representation.}
\label{subsec:blochrep} A two-dimensional quantum system, with state vector $ \ket{\psi} \in \mathbb{S}^3 \subset {\cal H}^2,$ is also called a {\em qubit}, representing the quantum mechanical ``unit'' of information in analogy with the classical ``bit'' \cite{nielsen-chuang}. 
The orthonormal reference basis elements  are typically denoted as $  \ket{0}, \,  \ket{1} $ and we have the vector representation  
\[
\ket{\psi} = c_0 \ket{0} + c_1  \ket{1} \simeq \begin{bmatrix} c_0 \\ c_1 \end{bmatrix},\] with $| c_0 |^2 + | c_1 |^2 =1,$ i.e. $\ket{\psi} \in \mathbb{S}^3.$
The observables for a two-level system form the 4-dimensional real vector space $\herm(\Hi^2)$. A convenient matrix basis for representing $\herm(\Hi^2)$ is given by the  {\em Pauli matrices}:
$$\sigma_x=
\begin{bmatrix}
0 & 1    \\
1 & 0    
\end{bmatrix},\quad \sigma_y=
\begin{bmatrix}
0 & -i    \\
i & 0    
\end{bmatrix},\quad
\sigma_z=
\begin{bmatrix}
1 & 0    \\
0 & -1    
\end{bmatrix},$$
completed with $\sigma_0=I_2$. $\sigma_{x,y,z}$ are Hermitian, traceless, unitary and involutive ($\sigma_j^2=I_2$).

In terms of the Pauli basis of $\herm(\Hi^2)$, the density operator of a two-level system can be expressed as 
\begin{equation}
\rho
= \tfrac{1}{2}(I_2 + x \sigma_x +y \sigma_y +z \sigma_z),
\label{eq:density-op-h2}
\end{equation}
where $ j = \tr{\rho \sigma_j} $, $ j=x,y,z$, are the expectation values along the observables $ \sigma_{x,y,z}$, and $ \um \tr{\rho I_2 } =1 $ is a constant (expectation value along the ``trivial'' observable $ \sigma_0 =I_2 $) ensuring
$ \tr{\rho} =1$. 
We define the {\em Bloch vector} as 
\[
\bm{\rho}= \begin{bmatrix}  x &  y &  z \end{bmatrix}^T \in \mathbb{R}^3.
\]
Due to the extra constant coordinate, the Bloch vectors are in fact affine, and form a convex subset of an hyperplane in $\mathbb{R}^{4} $.
The matrix (Hilbert-Schmidt) inner product on the space of Hermitian matrices induces the standard Euclidean inner product on $ \mathbb{R}^{4} $ (or $ \mathbb{R}^{3} $ for the linear part), indicated with $ \llangle  \, \cdot , \, \cdot \, \rrangle$. If $ {\rm e}_j $ is the canonical basis element in $ \mathbb{R}^{4} $: 
\[
\frac{\tr{\sigma_j \sigma_k }}{2} =  \delta_{j k}  \; \Longleftrightarrow \;  \llangle {\rm e}_j , \, {\rm e}_k \rrangle =\delta_{j k}, \quad \forall \; j, \, k = 0, \, x, \, y, \, z .
\]
Hence the {\em Frobenius norm} of $ \rho $ induces the standard Euclidean norm $ \| \, \cdot \, \| $ for $  \bm{\rho} $: $ \|  \bm{\rho} \| =  \sqrt{\tr{\rho^2}} $.
The condition $  {\rm  tr} (\rho^2 ) \leq 1 $ then translates in $ \bm{\rho} $-space as $  \bm{\rho} $ belonging to the solid ball of radius $ 1 $.
The surface $ \|  \bm{\rho} \| = 1$ of such ball corresponds to pure states. 
On the Bloch ball, the ground state $\ket{0}\bra{0}  $ corresponds to $ \bm{\rho} =[0,0,1]^T$, the excited state $\ket{1}\bra{1} $ to $ \bm{\rho} =[0,0,-1]^T$ and the completely mixed state $\frac{1}{2} \ket{0}\bra{0} + \frac{1}{2} \ket{1}\bra{1} $ to $  \bm{\rho} =[0,0,0]^T$.

\subsection{Joint systems and entanglement} 
Entanglement, its characterization, and its uses as a resource in communications and computing represent vast subfields in quantum information sciences: here we only aim to recall the basic definitions. Consider two quantum systems with associated Hilbert spaces $\Hi_A,\Hi_B$ respectively. The joint description is given in the tensor product space $\Hi_{AB}=\Hi_{A}\otimes\Hi_{B}$ (see Appendix \ref{notations}). Observables on one of the subsystems, say $X_A\in\herm(\Hi_A)$ are then mapped to $X=X_A\otimes I.$ However, any Hermitian operator on $\Hi_{AB}$ is a valid observable for the joint system.
It is easy to see that there exist state vectors $\ket{\psi}\in\Hi_{AB}$ that do not admit a factorized representation, yet represent perfectly valid states for the compound. These states are called {\em entangled}. Consider {\em e.g.} two qubits, with Hilbert spaces $\Hi_A,\Hi_B$. Then one can directly prove that \[\ket{\psi}=\frac{1}{\sqrt{2}}(\ket{0}\otimes\ket{0}+\ket{1}\otimes\ket{1})\neq \ket{\psi_A}\otimes\ket{\psi_B}\]
for any choice of $\ket{\psi_A}\in\Hi_A,\,\ket{\psi_B}\in\Hi_B.$ 
If the two systems are well-identified and physically separated, this implies that the joint system can be in principle be prepared in a pure state which is intrinsically non local.

In the density operator formalism, a state $\rho_{AB}\in\mathfrak{D}(\Hi_{AB})$ is usually said to be entangled if it cannot be written as a classical mixture of factorized density operators, i.e. if
\[\rho_{AB}\neq\sum_jp_j\rho^j_A\otimes \rho_B^j,\]
with $\sum_j p_j=1,\,p_j>0\,\forall j$ and $\rho_{A}^j\in\mathfrak{D}(\Hi_{A}),\;\rho^j_{B}\in\mathfrak{D}(\Hi_{B}).$

\section{Dynamical models of closed quantum systems}\label{sec:dynamics}

\subsection{The Schr{\"o}dinger equation}
The basic dynamical postulate of quantum dynamics is that the state vector $ \ket{\psi} $ of a closed system obeys the autonomous linear ODE
\beq
\begin{cases}
\hbar   \dot{\ket{\psi}}  & = -i H_0  \ket{\psi} , \qquad \ket{\psi} \in \mathbb{S}^{2N-1}  \\
\ket{\psi(0)} & = \ket{\psi_0},
\end{cases}
\label{eq:Schrod1}
\eeq
called the {\em Schr{\"o}dinger equation for the state vector}, where $H_0$ is the Hamiltonian of the system. Hereafter the units are chosen, as customary, so as to fix the Plank constant $ \hbar $ to 1.

The real eigenvalues $ \eneig_j $, $ j=1, \ldots , N $ of $ H_0 $ are  called the {\em energy levels} of the quantum system. Transformation of the energy levels by an additive constant only introduces in the state vector dynamics a global phase factor, which we already recognized to be physically irrelevant. 
To eliminate this ambiguity, it is customary to choose a traceless $ H_0 $: $ \eneig_1 + \ldots + \eneig_N =0 $.
We can assume that $ \eneig_j \leqslant  \eneig_k $ if $ j \leqslant k $, $ j, \, k \in  \left\{ 1, \ldots, N \right\}$.
The differences $  \eneig_j - \eneig_k $, $  j, \, k \in \left\{ 1, \ldots, N \right\} $, are called the {\em transition frequencies} (or {\em resonances}, or {\em Bohr frequencies}). A system is said to be {\em regular} (or {\em nondegenerate}) if $ \eneig_j \neq \eneig_k $ $ \forall \; j, \, k \in \left\{ 1, \ldots, N \right\} $, $ j \neq k $, i.e., if its energy levels are all nondegenerate.
A system is said {\em strongly regular} (or {\em with no degenerate transition}) if $ \eneig_j - \eneig_k \neq \eneig_\ell - \eneig_m $ $ \forall \; j, \, k, \, \ell, \, m \in \left\{ 1, \ldots, N \right\} $, $ j\neq k$, $\ell \neq m $, $ ( j,\,k ) \neq ( \ell , \, m )$, i.e., if its resonances are nondegenerate.

The control of a quantum mechanical system is typically obtained by coupling it with one or more tunable electromagnetic fields. Perturbation theory can be used to derive a model for this quantum - external field interaction: in many physically relevant situations, and reasonable approximations, \cite{alicki-lendi,dalessandro-book,petruccione}, a semiclassical description is sufficient, i.e., the series expansion can be truncated at the first order terms, yielding
\beq
\begin{cases}
\dot{\ket{\psi}}  & = -i (H_0 + \sum_j u_j H_j )   \ket{\psi}  , \qquad \ket{\psi} \in \mathbb{S}^{2N-1}  \\
\ket{\psi(0)} & = \ket{\psi_0},
\end{cases}
\label{eq:Schrod-multi}
\eeq
where the driving Hamiltonians $ H_j = H_j ^\dagger $ contain the couplings between the energy levels of the free Hamiltonian $ H_0 $, and $ u_j $, the amplitude of the interactions, represent our control parameters, $ u_j \in [u_{j,\rm{min}}, \; u_{j,\rm{max}}] \subset  \mathbb{R} $. A control entering the dynamics as in \eqref{eq:Schrod-multi} is sometimes called {\em coherent}, as it preserves the unitary evolution of the state vector (see below for a geometric explanation).
To avoid conflict with other meaning of the term ``coherent'' (see Section~\ref{sec:lin-feedb-net}), we shall call the controls of \eqref{eq:Schrod-multi} unitary. 
The system \eqref{eq:Schrod-multi} is a bilinear control system on a sphere \cite{Brockett3}.
We shall see that one control function is generically sufficient to ensure controllability, and most of the results we present generalize in a straightforward way to the multi control setting. Hence, we consider from now on the simple case:
\beq
\begin{cases}
\dot{\ket{\psi}} & = -i (H_0 + u H_1 )  \ket{\psi}  \\
\ket{\psi(0)} & = \ket{\psi_0}.
\end{cases}
\label{eq:Schrod2}
\eeq

\subsection{Unitary propagator}

The complex sphere $ \mathbb{S}^{2N-1}, $ representing pure states, is a homogeneous space of the Lie group $ U(N) = \{ U \in GL(N,\mathbb{C} ) \; |  \; UU^\dagger = U^\dagger U = I \} $ as well as of its proper subgroup $ SU(N) = U(N) / U(1) $, in which the global phase factor has been eliminated. 
The Schr{\"o}dinger equation can therefore be lifted to the Lie group $ SU(N)$, obtaining in correspondence of \eqref{eq:Schrod2} the right invariant matrix ODE:
\beq
\begin{cases}
\dot{U}  & = -i ( H_0 + u H_1 )  U , \qquad U \in SU(N)  \\
U(0) & = I,
\end{cases}
\label{eq:unitary_propag1}
\eeq
called the {\em Schr\"odinger equation for the unitary propagator}.
For the system \eqref{eq:unitary_propag1}, the total Hamiltonian $ H(t, u) = H_0 + u(t) H_1 $ is in general time-varying. 
The solution of \eqref{eq:unitary_propag1} is therefore given by a formal, time-ordered exponential 
\beq
U(t)= {\cal T} {\rm exp }\left(-i  \int_0^t H (s, u) ds \right),
\label{eq:dyson-series}
\eeq 
which is called Dyson's series in the physics literature and is analogous to a Volterra series in control theory.
Consequently, for \eqref{eq:Schrod2} we have: $ \ket{\psi(t)} = U(t) \ket{\psi_0} = {\cal  T} {\rm exp }\left(-i  \int_0^t H (s, u) ds \right)\ket{\psi_0} $.

The Lie algebras of $ U(N) $ and $ SU(N)$ are, respectively, $ \mathfrak{u}(N) = \{ A \in \mathbb{C}^{N \times N} \; | \;  A^\dagger = - A \} $ and $ \mathfrak{su}(N) = \{ A\in \mathfrak{u}(N) \; | \;  {\rm tr} (A) = 0 \} $. 
$ \mathfrak{u}(N) $ and $ \mathfrak{su}(N) $ are semisimple compact Lie algebras, meaning that the corresponding Killing forms are negative definite \cite{Jurdjevic1}.

\subsection{Quantum Liouville-von Neumann equation}
\label{subs:control-Liouv}
Given a certain Hamiltonian $ H(t, u)$, in this Section we describe the time evolution of the density operator which corresponds to the Schr{\"o}dinger equation of $ H(t,u)$. Since for any state vector $ \ket{\psi_j} $ we have $ \ket{\psi_j} = U(t) \ket{\psi_j (0) } $, for the outer product $ \ket{\psi_j}\bra{\psi_j} $ we obtain 
\[
\ket{\psi_j(t)}\bra{\psi_j(t) } =  U(t)\ket{\psi_j(0)}\bra{\psi_j(0) }  U^\dagger (t),
\]
and similarly for any convex sum of $ \ket{\psi_j}\bra{\psi_j} $.
In terms of the unitary propagator \eqref{eq:dyson-series} we thus have 
\beq
\rho(t) = U(t) \rho(0) U^\dagger (t).
\label{eq:liouville1}
\eeq
The infinitesimal version of \eqref{eq:liouville1} is the {\em quantum Liouville-von Neumann equation}:
\beq
\begin{cases}
\dot \rho & =  -i [ H (t,u) ,\, \rho ] \\
\rho(0) & =  \rho_0.
\end{cases}
\label{eq:liouville2}
\eeq
%
The main feature of this equation is that it generates {\em isospectral evolutions} i.e., 
\beq
{\rm sp} ( \rho(t)) =  {\rm sp} ( \rho(0))=  \Phi (\rho) =  \{ \mu_1, \ldots, \mu_N \} .
\label{eq:isosp-liouv1}
\eeq
A consequence of the isospectrality of \eqref{eq:liouville2} is that the eigenvalues $ \Phi(\rho)$ form a complete set of constants of motion of \eqref{eq:liouville2}. 
Hence $ \mathfrak{D}(\Hi) $ is foliated into (compact and connected) leaves uniquely determined by $ \Phi(\rho) $. 
Call $ \mathcal{C} \in  \mathfrak{D}(\Hi) $ one such leaf and consider $ \rho_0 \in \mathcal{C}$.
Then $ \mathcal{C} $ corresponds to the orbit of $ SU(N) $ under the conjugation action passing through $ \rho_0$: $ \mathcal{C} = \{ U \rho_0 U^\dagger ,\;  U \in SU(N) \}$.
If the geometric multiplicities of the eigenvalues $ \Phi( \rho_0 ) $ are given by $ j_1, \ldots, j_\ell$, $ j_1 + \ldots + j_\ell = N $, $ 2\leqslant \ell \leqslant N $, then $ \mathcal{C} $ is the homogeneous space
\[
\mathcal{C}  =  U(N) / \left( U(j_1) \times \ldots \times U(j_\ell )  \right) , \]
with $j_1 + \ldots + j_\ell = N ,\; 2\leqslant \ell \leqslant N.$ As we vary the eigenvalues $  \mu_1, \ldots, \mu_N$, the multiplicities $  j_1, \ldots, j_\ell$ form a flag; the $ \mathcal{C} $ are consequently called {\em complex flag manifolds}, see \cite{Bengtsson1}.
The flag determines also the (even) dimension of the manifolds $ \mathcal{C} $, call it $m$, which can vary from $ 2N-2 $ in the case of pure state $ \Phi = \{1, \, 0, \ldots , 0 \} $, to $N^2-N$ in the case of all different eigenvalues $ \Phi = \{ \mu_1, \mu_2, \ldots, \mu_N \} $, $ \mu_j \neq \mu_\ell $, $ \sum_{j=1}^N \mu_j =1 $.

Alternatively to $ \{ \mu_1, \ldots, \mu_N \} $, one can consider as a complete set of invariant quantities of the ODE \eqref{eq:liouville2} the so-called symmetric functions i.e., the coefficients of the characteristic polynomial $ {\rm det} (s I - \rho ) =0 $ or, equivalently, the quantities $ \tr{\rho}, \; \tr{\rho^2 } , \; \ldots , \tr{\rho^N } $.
See \cite{Lendi1} for a complete description of the invariants of motion.

\subsection{Schr{\"o}dinger's and Heisenberg's dual pictures of dynamics}

In the previous sections we studied the evolution of states of closed quantum systems, which for density operators turns out to be given by the conjugate action \eqref{eq:liouville1} of the unitary propagator $U(t)$ defined in \eqref{eq:dyson-series}. Assume the state is $\rho(t)=U(t)\rho_0U^\dag(t)$, and consider an observable $Y=\sum_jy_j\Pi_j$ so that
$\E_{\rho(t)}(Y)=\trace(Y\rho(t)).$
By using the cyclic property of the trace, however, we can also write
\[\E_{\rho(t)}(Y)=\trace(YU(t)\rho(0)U^\dag(t))=\trace(U^\dag(t)YU(t)\rho(0)),\]
so that by defining $Y(t)=U^\dag(t)Y(0)U(t),\,Y(0)=Y$, we get
$\E_{\rho(t)}(Y)=\E_{\rho(0)}(Y(t))$.
This shows how the same predictions on measurements can be obtained by assuming that {\em the state is time invariant}, and by letting the observable evolve according to conjugate action of the {\em adjoint} unitary operator. This corresponds exactly to the dual evolution with respect to the Hilbert-Schmidt product. Accordingly, the evolution equation for observables in this dual picture is
\[\dot Y=i[H(t,u),Y].
\]
Throughout most of the paper, dynamics will affect the system state while leaving the observable variables time-invariant (the so-called Schr\"odinger picture in the physics literature): we choose to adopt the view most commonly found in the existing literature. However, the dual (Heisenberg) picture is more natural at least in the framework of quantum probability, and we shall switch to it when describing quantum fields, quantum stochastic processes and feedback networks.

\subsection{Two-level system: Bloch equation}
\label{subsec:bloch_eq}
The foliated structure of $\mathfrak{D}(\Hi) $ is not apparent from the invariants $ \Phi(\rho)$. A way to obtain a better visualization, very effective in the case $N=2 $, is to use the Bloch vector parametrization.
Any traceless Hamiltonian $ H(t) $ can be decomposed in terms of the Pauli matrices as $ H (t) =  \sum_{j=x,y,z} h_j (t)\sigma_j  $.
From \eqref{eq:density-op-h2}, also $ \rho $ can be expressed in terms of the $ \sigma_j$ matrices.
Consider the Liouville equation \eqref{eq:liouville2}. 
Using the Bloch vector $ \bm{ \rho} $, one gets along $ -i \sigma_j $:
\beq
- i [ \sigma_j , \rho ] \simeq B_j \bm{ \rho} ,
\label{eq:adj_rep1}
\eeq
where $ ( B_j )_{ k \ell  } =  c_{j \ell }^k $ are the $ 3 \times 3 $ skew-symmetric matrices of structure constants associated to the $ \{ -i \sigma_{x,y,z} \} $ basis of $\mathfrak{su}(2)$.
Identifying as in \eqref{eq:adj_rep1} a matrix commutator and a matrix-vector multiplication through the structure constants means passing to the adjoint representation of a Lie algebra.
For $\mathfrak{su}(2)$, the adjoint representation corresponds to the Lie algebra $\mathfrak{so}(3)$. 
With respect to $ \{ -i \sigma_{x,y,z} \} $, a basis of $\mathfrak{so}(3)$ is given by the following $  3 \times 3 $ skew-symmetric matrices:
\beq
\!\! B_x \!  =2 \!\begin{bmatrix} 0\! & -1\! & 0 \!\\ 1\! & 0\! & 0\! \\ 0 \!& 0\! &\! 0 \end{bmatrix}\!\!,  
B_y \!  = 2 \!\begin{bmatrix} 0\! & 0\!  & 1\! \\ 0\! & 0\! & 0\! \\ -1\! & 0\!  & 0\! \end{bmatrix}\!\!,   
B_z \!  =2 \!\begin{bmatrix} 0\! & 0\! & 0\! \\ 0 \!& 0\! & -1\!  \\ 0 \!& 1\! & 0\! \end{bmatrix}\!.
\label{eq:so3-basis}
\eeq
Hence
\[
\! - i [ H , \rho ]  = -i\sum_{j=x,y,z} [ h_j \sigma_j,  \rho] 
 \simeq \sum_{j=x,y,z}  h_j B_j \bm{ \rho} = B_H \bm{ \rho}
\]
and the Liouville - von Neumann matrix ODE \eqref{eq:liouville2} becomes the following vector ODE, normally called the Bloch equation:
\beq
\dot{\bm{\rho}} = B_H  \bm{\rho} =
\sum_{j=x,y,z}  h_j B_j \bm{ \rho}.
\label{eq:liouville3}
\eeq
The Bloch equation \eqref{eq:liouville3} describes infinitesimal rotations of $ \bm{\rho} $. 
As such, $ \| \bm{\rho} \| $ is a constant of motion. 
The time evolution of \eqref{eq:liouville3} occurs therefore on a sphere of radius $ \| \bm{\rho} \|= \| \bm{\rho}(0) \| $, and indeed $ \tr{\rho^2} $ is an invariant of motion.

\subsection{$N$-level system: explicit parametrizations}

For a $N$-level system, the density operator has $ N^2 -1 $ degrees of freedom.
Several explicit parametrizations are used in the literature to ``vectorize'' the matrix ODE \eqref{eq:liouville2}:
\begin{itemize}
\item coherence vector, i.e., vector of expectation values with respect to a complete set of orthonormal traceless Hermitian matrices ({\em e.g.} the $N$-dimensional Pauli matrices), see \cite{alicki-lendi};
\item superoperator representation, where the matrix state space is transformed into vector state space {\em e.g.} through the ``vec'' operation. This is refered to as {\em Liouville space} representation in the literature, see \cite{Ernst1}.
\end{itemize}

The explicit description of $ \rho$ in terms of these parametrizations is however much more complex than for the two-level case, see for example \cite{Schirmer7}.


\section{Dynamical models for open quantum systems}\label{sec:open}

\subsection{Hamiltonian description of open quantum systems}\label{subsec:opensys}

Any model for a physical, realistic quantum system should in general take into account the interactions of the system of interest with its environment. The most general way to introduce open-system dynamics is via its {\em Hamiltonian description}: consider a finite-dimensional quantum system $\Si$, coupled to a quantum environment
$\Ei$ via some interaction Hamiltonian. We can in principle think of including in the environment any other system that interacts with the one of interest, making the compound system isolated. 
More precisely, let $\Hi_\Si$ and $\Hi_\Ei$ denote the system and the environment Hilbert spaces, with $ \dim(\Hi_\Si)= N<\infty$. Assume that the dynamics of the joint system plus environment is driven by a time-independent Hamiltonian
of the form \beqa\label{hamiltonianopen} H_{tot}=H_\Si\otimes
I_{\Ei}+I_\Si\otimes H_\Ei+ H_{\Si\Ei},
\eeqa
\noindent
where $H_\Si,H_\Ei,H_{\Si\Ei}$ are the
system, environment, and interaction Hamiltonians,
respectively. 
On the joint Hilbert space $\Hi_\Si\otimes\Hi_\Ei$ the dynamics is still unitary, and could be described along the lines of the former sections. 

\subsection{Reduced states and the master equation approach}

In many situations, {\em e.g.} in order to limit the complexity of the description or because the degrees of freedom of the environment are not accessible, it is convenient to seek a reduced description, involving only the system $\Si$. This can be formally obtained by ``averaging'' over the environment degrees of freedom via the partial trace \cite{alicki-lendi} (see Appendix).
Assume the initial state for the system-environment ODE is factorized, $\rho_0\otimes\rho_\Ei$, and define $U_{\Si\Ei}(t)=\exp(-iH_{tot}t)$;
then one can define:
\[\rho(t)={\cal
T}_{(t,0)}[\rho_0]=\trace_E\left(U_{\Si\Ei}(t)(\rho_0\otimes\rho_\Ei) U_{\Si\Ei}^\dag(t)\right).\]
The resulting dynamics is in general non-Markovian, and an integro-differential equation for the reduced state can be formally derived  by using the Nakajima-Zwanzig projection technique \cite{petruccione}, which typically involves a non-trivial memory kernel, and can be written in the form: 
\begin{equation}\label{NZ}
\frac{d}{dt}\, \rho(t) = \int_{t_0}^t \mathcal{K}(t, s)\rho(s)\,
ds\ , \ \ \ \rho(t_0)=\rho_0 .
\end{equation}

While control of non-Markovian dynamical systems has been considered, especially in the Hamiltonian picture (see Section \ref{decoupling}), most of the control-oriented literature has been concerned with the Markovian case, which has the enormous advantage of leading to dynamical equations in simple form.
In fact, under proper assumptions on the memory time-scales of the environment, or by invoking the Born-Markov approximation or another appropriate limiting procedure \cite{alicki-lendi,petruccione,wiseman-book}, in certain situations the family $\{{\cal
T}_{(t,s)}[\cdot],\,t,s\geq0\}$ becomes actually time-homogeneous ($\forall \tau,$ ${\cal
T}_{(t,s)}={\cal
T}_{(t+\tau,s+\tau)}=:{\cal
T}_{t-s}$) and satisfies the following properties: 
\begin{enumerate}
\item  ${\cal T}_t$ is linear and continuous for all $t\geq 0,$ and maps ${\mathfrak D}(\Hi_S)$ into itself;

\item ${\cal T}_t$ is completely positive \cite{alicki-lendi,nielsen-chuang}, that is, $\forall\,t\geq 0,$ ${\cal T}_t(\rho)=\sum_k M_{t,k}\rho M_{t,k}^\dag,$
for some finite set of operators $\{ M_{t,k}\}\subset\bound(\Hi_S)$;
\item the forward composition law holds: ${\cal T}_t\circ{\cal T}_s={\cal T}_{t+s}$. 
\end{enumerate}
1)-2) are physically motivated (and hold for general ${\cal
T}_{(t,s)}$), while 3) is satisfied in a significant variety of situations, {\em e.g.} in many quantum optical systems \cite{wiseman-milburn,carmicheal} or whenever the memory effect of the environment can be neglected \cite{petruccione}. In this cases the state dynamics are described by a Markovian {\em Quantum Dynamical Semigroup} (QDS) \cite{alicki-lendi}. It has been proved \cite{lindblad,gorini-k-s} that the generator for a QDS is a {\em Markovian Master Equation} (MME) of the  form: 
\beqa
&&\! \! \!  \ddt \rho(t)=\mathcal{L}(\rho(t)) = -{i}[H,\rho(t)]+\mathcal{L}_D(\rho(t)) 
\label{eq:master1} \\
&&\! \! \!  =-{i}[H,\rho(t)]+\sum_{k,l=1}^{N^2-1} a_{kl}\left(\lambda_k\rho(t)\lambda_l^\dag-\frac{1}{2}\{\lambda_k^\dag
\lambda_l,\,\rho(t)\}\right), \nonumber
\eeqa
with $\{ \lambda_j\}_{j= 1,\ldots, N^2-1} $ a complete orthonormal set of traceless $ N\times N $ Hermitian matrices, for example the $ N$-dimensional Pauli matrices, or the so-called Gell-Mann matrices \cite{alicki-lendi}. The effective Hamiltonian $H$ is in general equal to the free system Hamiltonian $H_{\cal S}$ plus some correction induced by the interaction with the environment (Lamb shift).  The positive semi-definite $A=(a_{kl})$ is called GKS (Gorini-Kossakowski-Sudarshan) matrix and fully specifies the dissipative part of the generator weighting the non-Hamiltonian terms $\lambda_k\rho(t)\lambda_l^\dag-\frac{1}{2}\{\lambda_k^\dag
\lambda_l,\,\rho(t)\}$. The MME \eqref{eq:master1} is a linear (or bilinear when control is present) matrix ODE, and the resulting dynamics is in general non-unitary.  A generator \eqref{eq:master1} is said to be {\em unital} when $ \mathcal{L} (I_N) = 0 $, and hence the completely mixed state $I_N/N$ is stationary. 
%
Thanks to the Hermitian character of $A$, by choosing a different operator basis, Eq. \eqref{eq:master1} can be put in a symmetrized form, also known as {\em Lindblad form} or Lindblad master equation:
\beqa
\ddt \rho(t)& = & -{i}[H,\rho(t)]+\hspace{-1mm}\sum_{k} {\cal D}(L_k,\rho(t))
\label{eq:master2} \\
& = &-{i}[H,\rho(t)]+\hspace{-1mm}\sum_{k} L_k\rho(t)L_k^\dag-\frac{1}{2}\{L_k^\dag L_k,\,\rho(t)\}. \nonumber
\eeqa
In many situations, it is reasonable (and convenient) to assume a MME in Lindblad form on a phenomenological basis \cite{alicki-lendi}. The $L_k$s are also called noise operators, or noise channels. In this way, however, the separation between the Hamiltonian and the dissipative part is not unique due to the loss of the trace and orthogonality constraints on the $L_k$s.
\subsection{Stability of the MME}
\label{subsec:stability}
While unitary dynamics necessarily correspond to generators with eigenvalues on the imaginary axis, the non-unitary terms in the MME introduce eigenvalues in the left complex half-plane. This motivates the study of the stability of the autonomous MME, a necessary first step towards the development of control strategies aiming at the preparation of desired states. 

Let us briefly recall some basic definitions and explore some stability features of the class of MME under scrutiny.
%
A set ${\cal M}$ is {\em invariant} for \eqref{eq:master2} if for any $\rho_0\in{\cal M},$ its trajectory ${\cal T}_t(\rho_0)\in{\cal M}$  for all $t\geq 0.$ Define a distance between a state and a set by
$
d({\rho},{{\cal M}}):=\inf_{\sigma\in{\cal M}}\|\rho-\sigma\|_1,
$ 
where $\| X \|_1=\trace(\sqrt{X^\dag X})$ is the trace norm.
Then we say that ${\cal M}$ is (marginally) {\em stable} if it is invariant and for every $\varepsilon>0$ there exists $\delta$ such that if $d(\rho_0,{\cal M})\leq \delta$ then $d({\cal T}_t(\rho_0),{\cal M})\leq \varepsilon$ for all $t\geq 0;$ ${\cal M}$ is {\em Globally Asymptotically Stable} (GAS) if it is stable and 
for any initial condition $\rho_0\in{\mathfrak D}(\Hi),$ $\lim_{t\rightarrow +\infty} d({\cal T}_t(\rho_0),{\cal M})= 0.$

In our case, while $\Li$ is indeed linear in $\rho$, the state manifold is the convex, compact set $ {\mathfrak D}(\Hi)$, hence overall we have a nonlinear system. Since ${\mathfrak D}(\Hi)$ contains a basis for the space of Hermitian operators, the action of ${\cal L}$ can then be extended (by linearity) to the real vector space $\herm(\Hi) .$
Thus, any matricial representation of such a generator on $\herm{}$ over the real field admits a real Jordan canonical form. This implies that it admits either real or
pairs of complex-conjugate eigenvalues. 

Since $\Li$ generates a semigroup of contractions in trace norm, $\{ {\cal T}_t=e^{\Li t} \}$ \cite{alicki-lendi,nielsen-chuang}, its eigenvalues must lay in the left complex half-plane. This implies that if a set is invariant, then it is also marginally stable. 
We shall thus focus on global asymptotic stability of equilibria. Since ${\cal T}_t$ maps the compact set ${\mathfrak D}(\Hi)$ into itself, by Brower's fixed point theorem \cite{ziedler-functional}, the semigroup must admit at least one invariant state $\rho_e\in{\mathfrak D}(\Hi),$ and $\rho_e$ must be in the kernel of $\Li.$
The following Theorem from \cite{schirmer-markovian} ensures that there do not exist isolated centers.
\begin{theorem}
\label{unique-attractive} 
A steady state of \eqref{eq:master2} is GAS if and only if it is unique.
\end{theorem}

In many cases, however, the invariant sets are more complex. The case of invariant subsystems has been treated in \cite{ticozzi-markovian,ticozzi-QDS}, while a decomposition of the Hilbert space that highlights the structure of attractive sets has been derived in \cite{baumgartner-2}. We recall here a linear-algebraic approach to the stability of subspaces.  Consider a decomposition of the Hilbert space $\Hi := \Hi_S \oplus
\Hi_R.$  Let $\{\ket{s_i}\}$ and
$\{\ket{r_j}\}$ be orthonormal bases for $\Hi_S$ and $\Hi_R,$
respectively.  The (ordered) basis
$\{\ket{s_1}, \ldots \ket{s_m}, \ket{r_1},\ldots,\ket{r_{n-m}}\}$
induces the following block structure on the matrix representation of
an arbitrary operator $X \in\mathfrak{B}(\Hi)$:
\begin{equation} 
\label{eq-block-structure}
X = \begin{bmatrix}X_S & X_P \\ X_Q & X_R \end{bmatrix}.
\end{equation}
Let us define the support of an operator as $\supp(X):= \ker(X)^\bot,$ and the set of states with support on $\Hi_S$ as:
$${\cal M}_S=\Big\{\rho\in\mathfrak{D}(\Hi)\,|\,\rho=\begin{bmatrix}\rho_S
& 0 \\ 0 & 0\end{bmatrix}, \:\rho_S\in\mathfrak{D}(\Hi_S)\Big\},$$
\[\forall\rho_0\in{\cal M}_S,\;t>0,\;{\cal T}_t(\rho_0)\in{\cal M}_S.\]
As a first step, we would like to characterize the generators that render a certain subspace invariant for the evolution.
\begin{lemma}\label{lemmainvariance}
Assume that $\Hi=\Hi_S\oplus\Hi_R$, and let $H$, $\{L_k\}$ be the
Hamiltonian and the noise operators in
\eqref{eq:master2}. Then, ${\cal M}_S$ is invariant {\em iff}
$\forall k$
\begin{equation}
\begin{split}
&L_k=\left(\begin{array}{c|c}
L_{S,k}&L_{P,k}\\
\hline
0&L_{R,k}
\end{array}\right),\\
&iH_P-\frac{1}{2}\sum_k L_{S,k}^\dagger L_{P,k}=0.
\end{split}
\end{equation}
\end{lemma}
Given that an invariant set is automatically marginally stable, it is possible to show (via LaSalle invariance principle) that ${\cal M}_S$ is GAS if and only if there is no invariant subspace with support on $\Hi_R.$ Refined or slightly different formulations of this result have been derived in \cite{ticozzi-markovian,schirmer-markovian}.

\subsection{Two-level MME in Bloch vector representation}\label{openbloch}

While the commutator (Lie bracket) is linear for the basis  elements $ \sigma_{x, y,z} $, the anticommutator has an affine structure: $
\{ \sigma_j , \, \sigma_k \} = \beta \delta_{j k} I + \sum_{\ell=1}^3 d_{j k}^\ell \sigma_\ell$, with $ d_{j k}^\ell $ a real and fully symmetric tensor and $ \beta $ a normalization constant.
From this property, it follows that in terms of the Bloch vector ${\bm \rho}$ the non-unitary part of \eqref{eq:master1} is in general an affine vector field.
Therefore the MME \eqref{eq:master1} becomes an inhomogeneous vector ODE 
\beq
\dot{\bm{\rho}}  = B_{H} \bm{\rho} + \sum_{j,k=1}^{3} a_{jk} \left( \Gamma_{jk} \bm{\rho} + g_{jk} \right),
\label{eq:lindblad-bloch}
\eeq
where the $ 3\times 3 $ symmetric matrices $ \Gamma_{jk} $ and the vectors $ g_{jk} $ are functions of the structure constants $ c_{\ast \ast}^\ast $ and $ d_{\ast \ast}^\ast$, see \cite{Cla-contr-open1} for details.
The vectors $ g_{jk} $ constitute the inhomogeneous part of the Lindbladian action.
When $ g_{jk}\neq 0 $ for some $ j, k$, then the origin of the Bloch ball cannot be a fixed point. 
Hence in terms of $ \bm{\rho}$, unital Lidbladians correspond to linear (instead of affine) dissipation generators.

\begin{example} [Two-level system with unital dissipation]
Assume the $ g_{jk} =0 $ in \eqref{eq:lindblad-bloch} and call $ \Gamma = \sum_{j,k=1}^{3} a_{jk} \Gamma_{jk} $. 
As soon as $ \Gamma $  is invertible, the only fixed point admissible in \eqref{eq:lindblad-bloch} is the origin of the Bloch ball, $ \bm{\rho} =0 $, i.e., the completely mixed state $ \rho = \frac{1}{2} I_2 $, and from Theorem \ref{unique-attractive} it is GAS. 
A common example for unital dissipation is the following 
\beq
\Gamma = - \begin{bmatrix} \gamma_1 \\ &  \gamma_2  \\ & & \gamma_3 \end{bmatrix},  \qquad  g =  0
\label{eq:example-unital}
\eeq
with $ \gamma_i \geqslant 0 $.
Several standard (discrete-time) noise channels used in quantum information, such as the bit flip, bit-phase flip and the phase flip, can be described in continuous-time by infinitesimal generators of the type \eqref{eq:example-unital}. 
For example, the phase flip channel squeezes the Bloch sphere along the $ x $ and $ y $ axis, leaving $ z$ unchanged: { the equivalent effect in continuous time, or {\em dephasing},} can be generated by \eqref{eq:example-unital} with $ \gamma_1, \, \gamma_2 \neq 0$, $ \gamma_3 =0$.
In this case the entire $ z $ axis of the Bloch ball is a set of fixed points for the system when $ u=0$. 

\end{example}

\begin{example} [Two-level atom with decay] Consider a two-level atom, with ground and excited states denoted by $\ket{0},\,\ket{1},$ respectively. Assume there is an average, observed rate of decay from the excited to the ground state equal to $\gamma$, i.e., if the state is initially in $\ket{1},$ the probability of finding it in the same state at time $t$ is $e^{-\gamma t}$. Assuming  Markovian dynamics and using the ladder operators $ \sigma_\pm = \um (\sigma_x \pm i \sigma_y) $, one is led to a generator of the form (see {\em e.g.} \cite{nielsen-chuang}, Sec. 8.4.1):
\beq
\dot{\rho} = -i [\Delta\sigma_z,\rho] + {\gamma} \left( \sigma_+ \rho \sigma_-  - \frac{1}{2}\{ \sigma_- \sigma_+,\rho\}\right),
\label{eq:sp-em}
\eeq

\noindent with $\Delta\sigma_z$ the system Hamiltonian (with trace normalized to zero), and $\sigma_+$  the noise operator associated to the decay channel ($ \sigma_- =\sigma_+^\dag$). In fact, if one computes the probability $p_1(t)=\trace(\Pi_1\rho(t)),$
with $\Pi_1=\ket{1}\bra{1}$, with initial state $\rho(t)=\Pi_1,$ indeed one gets $p_1(t)=e^{-\gamma t}$.
In terms of \eqref{eq:lindblad-bloch}:
\beq
\Gamma = - \gamma\begin{bmatrix} \frac{1}{2} \\ &  \frac{1}{2}  \\ & & 1 \end{bmatrix},  \qquad  g =  \gamma\begin{bmatrix} 0 \\ 0 \\ 1 \end{bmatrix}.
\label{eq:example-affine}
\eeq
\end{example}

\subsection{Quantum stochastic models}
For open-loop control, we introduced a controlled Schr\"odinger equation  in which the field action is modeled as a semiclassical perturbation of the system Hamiltonian. Performing measurements on light fields that have interacted with a system is also a common way of acquiring information about its state: however, to model this we shall resort to a fully quantum description of the Electro-Magnetic (EM) field.
%
%
The next subsections will sketch the key ingredients of a model for this class of open systems. We anticipate that, while not every quantum control system is properly (or conveniently) described in this framework, these models present at least two advantages:
(i) Under suitable (Markovian) assumptions, the interaction presents a natural input-output structure, that lends itself to the study and design of interconnections, with localized systems, beam splitters and optical elements as blocks, and traveling fields as signals; and (ii) the same formalism can be used for both the case of monitored outgoing fields, and the case of a unitarily evolving quantum network.


These topics can be approached from many directions, and using different mathematical tools: see e.g. \cite{wiseman-feedback,transferfunction-1,gardiner-qn} for thorough treatment of the field input-output formalism, 
\cite{Barc09,wiseman-book} for models and applications to quantum measurements of light, \cite{parthasarathy,maassen-qp} for a quantum probabilistic approach to quantum stochastic processes, \cite{vanhandel-filtering,belavkin-filtering} for derivation of the quantum filters. 





\subsubsection{Electromagnetic field and quantum stochastic processes}

A general quantized EM field traveling along the $z$ axis can be thought as a collection of quantum harmonic oscillators, each of these corresponding to a {\em mode} of the field at a given frequency, see Appendix~\ref{qosc}.
Under proper assumptions, the quantum vector potential and the electric field propagating along $z,$ and with frequency content concentrated at a given $\omega,$ can be described as linear combinations of $b(z,t)$ and $b^\dag(z,t),$ namely the annihilation and creation operators for the mode of frequency $\omega$ at position $z$ and time $t.$ These field operators satisfy the commutation relation $[b(z,t),b^\dag(z,t')]=\delta(t-t')I$, where $\delta$ is the Dirac delta function, generalizing the harmonic-oscillator case to a traveling field. The operator valued processes $b(z,t),b^\dag(z,t)$ are thus remarkably singular, and they play the role of stochastic processes in the quantum domain \cite{parthasarathy,vanhandel-filtering}. Let us skip what would be a rather technical detour, and just mention that a proper white noise theory and It$\hat{\mbox{o}}$ calculus can be developed: we will recall here the basic definitions and rules on the manipulations of these operator valued processes.

We shall first focus on the description of the field right before it interacts with the system, that we assumed localized at $z=0$, and hence at $z=0^-$. Define
\[
B_{in}(t)=\int_{0}^{t}b(0^-,s)ds,\quad
\Lambda_{in}(t)=\int_{0}^{t}b^\dag(0^-,s)b(0^-,s)ds.
\]
It can be shown that $d\Lambda_{in}(t)$ is the quantum equivalent of a classical Poisson process, and it is diagonal in the number basis $\{\ket{\psi_n}\}$. Furthermore, the self adjoint {\em field quadratures} $dB_{in}+dB_{in}^\dag,$ $-i(dB_{in}-dB_{in}^\dag)$ both correspond (in their diagonal basis) to classical Wiener processes, but do not commute. These operators act on a (infinite-dimensional) continuous tensor product of Hilbert spaces, called the Fock space \cite{gardiner-qn}, denoted $\Hi_F$.
Assume the input field is in the {\em vacuum state}, that is $\rho_0=\ket{\psi_0}\bra{\psi_0},$ where $\ket{\psi_0}$ is the zero eigenvalue of $\Lambda_{in}.$ The following multiplication rules (non-commutative It$\hat{\mbox{o}}$ table) can be used for calculus with the fundamental processes we just introduced:
\beqan
& dB_{in}(t)dB_{in}^\dag(t)=I dt, & \\
&  dB_{in}^\dag(t)dB_{in}(t)=dB_{in}(t)dB_{in}(t)=dB_{in}^\dag(t)dB_{in}^\dag(t)=0,& \\
& d\Lambda_{in}(t)d\Lambda_{in}(t)=d\Lambda_{in}(t), & \\
& d\Lambda_{in}(t)dB^\dag_{in}(t)=dB^\dag_{in}(t),\; dB_{in}(t)d\Lambda_{in}(t)=dB_{in}(t).&
\eeqan
These rules must be modified in order to apply to {\em different input field states} \cite{transferfunction-1}. It is worth remarking that a stochastic calculus has also been developed for fermion fields \cite{hudson-fermions}. These non-commutative stochastic processes are the cornerstones for constructing the model of the system-field joint dynamics, which we shall introduce next.

\subsubsection{Joint evolution and Heisenberg dynamics}
We now have all the ingredients to describe the dynamical interaction of the free field with the system. The joint dynamics on $\Hi_S\otimes\Hi_F$, assuming that no other interaction affects either subsystem, is unitary and in general governed by a Quantum Stochastic Differential Equation (QSDE) of the form:
\beqa dU(t)&=&\Big((S-I)\otimes d\Lambda_{in}(t)+L\otimes dB_{in}^\dag(t)\label{QSDE}\\&&-L^\dag S\otimes dB_{in}(t)-((iH+\um L^\dag L)\otimes I) dt\Big)U(t), \nonumber
\eeqa
with $U(0)=I,$ $H$ the free Hamiltonian of the system, and $L,S$ in our setting are bounded system operators specifying the interaction of the system with the field, with $S$ unitary.

In terms of the solution $U(t)$ of \eqref{QSDE} one can now find 
the Heisenberg-picture dynamics for the system operators and the field processes. For an initial condition $X(0)=X_0\otimes I,$ the evolution is simply 
$X(t)=U^\dag(t)X(0)U(t),$ and the {\em output} field operator is $B_{out}(t)=U^\dag(t)I\otimes B_{in}(t) U(t).$
Using this notation also for the $S,L$ operators, the stochastic differential equation for the input-output relation takes the simple form:
\beq\label{inputoutput} dB_{out}(t)=S(t)dB_{in}(t)+L(t)dt. \eeq

%

\subsubsection{Conditional expectation, homodyne detection measurement and filtering equations}

From the framework we outlined above, it is possible to rigorously derive equations for the system operators and state dynamics conditioned on the measurement outcome. To begin with, we need to introduce quantum conditional expectations. Consider a state $\rho,$ an operator $X(t),$ and a self-adjoint family (process) $\{Y(s)\}_{s\in[0,t]}\}$ on $\Hi_S\otimes\Hi_F,$ such that $[Y(r),Y(s)]=0$ for all $r,s\in [0,t].$ 
$\{Y(t)\}$ is then called {\em self-nondemolition}, and it generates a commutative (von Neumann) operator algebra ${\cal Y}$ \cite{vanhandel-filtering}. 
Let ${\cal Y}'$ denote the set of observables that commute with any $Y\in{\cal Y},$ that is the {commutant} of ${\cal Y}$. A {\em conditional expectation} of $X(t)$ with respect to the observations of a process $Y(t)$ up to time $t$,
is defined as an operator $\hat X(t)=\pi_t(X)\in {\cal Y}'$ such that
for all $Z\in {\cal Y}'$ 
$\E_\rho(\hat X(t)Z)=\E_\rho(X(t)Z).$
This is equivalent to a least-square estimation of the operator, and can be interpreted geometrically as the orthogonal projection of $X$ onto ${\cal Y}$ \cite{vanhandel-filtering}.

Let us illustrate the key steps of the derivation of the general filtering equation in a physically relevant case. Assume in fact we are monitoring the quadrature process 
$Y(t)=B_{out}(t)+B^\dag_{out}(t)$ and that for our system $S=I$. By using \eqref{inputoutput}, we have that $Y(t)$ has  differential
\beq dY(t)=(L(t)+L^\dag(t))dt + dB_{in}(t)+dB^\dag_{in}(t).\label{measproc}\eeq
In this setting, the differential for a general system operator takes the form
\[dX(t)={\cal L}^\dag(X(t)) dt+dB_{in}^\dag [X(t), L(t)] +[L^\dag(t), X(t)]dB_{in}(t),\]
where
${\cal L}^\dag (X)=i[H,X]+L^\dag XL-\um\{L^\dag L,X\}$
is the adjoint of a MME generator in Lindblad form.
This process can be used as a (at least approximate) model for experimental homodyne detection of the outgoing field. We also assume, for the sake of simplicity, that the photodetectors employed in the monitoring have efficiency one, but the models can be easily adapted to limited detection capability.
It is easy to show that $Y(t)$ satisfies the two properties:  (i) $[Y(t),Y(s)]=0$ at all times $s,t$, and (ii) $[X(t),Y(s)]=0$ for all the observables on the systems $X(t)=U(t)^\dag (X_0\otimes I) U(t).$ Property (i) ensures that the output signal is equivalent to a classical stochastic process, and with property (ii) it also guarantees that the conditional expectation for any system operators does exist.
Thus, it is possible to derive a stochastic differential equation for the conditional expectation, namely the {\em quantum filter} \cite{belavkin-filtering};
\beqa d\pi_t(X) &=& \pi_t({\cal L}^\dag (X))dt\label{belavkinfilter}\\
&&+\Big(\pi_t(XL + L^\dag X)-\pi_t(L+L^\dag)\pi_t(X)\Big)dW(t),\nonumber\eeqa
where $dW(t)=dY(t)-\pi_t(L+L^\dag)dt$ is the {\em innovation process}. 
It can be shown by the spectral theorem (see e.g. \cite{vanhandel-filtering}) that for filters in this form the operator-valued innovation process can be substituted by a real-valued, classical Wiener process.

Conditional dynamics in the Schr\"odinger picture can be derived by defining the conditional state $\rho_t\in{\mathfrak D}(\Hi_S)$, as the system density operator such that $\E_{\rho_t}(X)=\E_\rho(\hat X(t)).$
From this relation and \eqref{belavkinfilter}, one can derive  the quantum filtering or
Stochastic Master Equation (SME) \cite{belavkin-filtering} in It$\hat{\mbox{o}}$ form:
\begin{equation}
d\rho_t=
\left( -i[H,\rho]
+{\cal D}(L,\rho_t)\right) dt
+ 
{\cal G}(L,\rho_t) d W_t, \label{eq:SME}
\end{equation}
where
\begin{equation}
\begin{split}
&\hspace{-9mm}{\cal D}(L,\rho)=L\rho L^\dag-\frac{1}{2}(L^\dag L\rho+\rho L^\dag L),\\
&\hspace{-9mm}{\cal G}(L,\rho)=
(L\rho+\rho L^\dag-\tr{ (L+L^\dag) \rho} \rho)
\end{split}
\end{equation}
are respectively the drift and diffusion parts introduced by the measurement of the field quadrature.
Here $ \{\rho_t\} $, the solution of
\eqref{eq:SME} given an initial condition $\rho_0$, exists, is unique, adapted to the filtration $\mathfrak{E}_t$ generated by the (classical) white noise $W_t,$ and it is $
\mathfrak{D}(\Hi)$-invariant by construction, see
\cite{belavkin-filtering,vanhandel-feedback}.
It is indeed possible to rewrite \eqref{eq:SME} in terms of the coherence vector, obtaining a standard non-linear diffusion.

\begin{example} [Filtering equation for a two-level system]\label{examplestoch}
Consider a 2-level system interacting with an electromagnetic field along the $z$-axis \cite{mirrahimi-stabilization}. Assume the system Hamiltonian is $H(t)= \Delta \sigma_{z}$, where $\Delta$
is a real parameter, and that the only interaction of the system is with a traveling field as described above.
The outgoing field is monitored via homodyne detection, and the corresponding SME is 
\begin{equation}\label{SME3}\begin{split}
d\rho_t=&-i [\Delta\sigma_{z}, \rho_t]dt+\nu(\sigma_{z}\rho_t\sigma_z- 
\rho_t)dt\\&+\sqrt{\nu}(\sigma_{z}\rho_t+\rho_t \sigma_{z}-2\tr{
\sigma_{z}\rho_t)\rho_t}dW_t,
\end{split}
\end{equation}
where $\nu$ is a parameter representing the measurement strength,
and $dW_t$ is the innovation process associated to the measurement record
\begin{equation}\label{Wiener2}
dW_t=dY_t-2\text{tr}(\sigma_{z}\rho_t)dt .
\end{equation}
This equation is of the form \eqref{eq:SME} with $H=\Delta\sigma_z,$ $L=\sqrt{\nu}\sigma_z.$
If we rewrite the dynamics in terms of the Bloch vector as in \eqref{eq:density-op-h2}, we get: 
\begin{equation}
\label{eq:bloch1}
\begin{split}
dx_t &= (-\Delta y_t-\tfrac{1}{2}\nu x_t)\,dt-\sqrt{\nu}\,x_tz_t\,dW_t \\
dy_t &= (\Delta x_t -\tfrac{1}{2}\nu y_t)\,dt-\sqrt{\nu}\,y_tz_t\,dW_t \\
dz_t &= \sqrt{\nu}\,(1-z_t^2)\,dW_t.
\end{split}
\end{equation}
This is a nonlinear diffusion in the closed unit ball of $\R^3.$ It is easy to verify that there are two equilibrium points for the uncontrolled dynamics: for both $[x,y,z]^T=[0,0,1]^T,$ and $[x,y,z]^T=[0,0,-1]^T$, we have $dx=dy=dz=0.$ These states correspond to the eigenstates of the measurement operator $\sigma_z.$
By direct computation \cite{vanhandel-feedback} it is possible to show that the {\em variance} of the observable $\sigma_z,$ $V_t(\rho_t)=\trace(\sigma_z^2\rho_t)-\trace^2(\sigma_z\rho_t)\geq 0,$ represents a stochastic Lyapunov function for the system in the sense of \cite{kushner}, and that all the trajectories converge in probability to the equilibrium states. From a physics viewpoint, the free dynamics of a system like \eqref{eq:bloch1} asymptotically replicate the {\em effect of a projective measurement}, with the correct probabilities of ``collapsing'' on the eigenstates of $\sigma_z$ induced by the initial state \cite{adler,vanhandel-feedback}.

\end{example}

\begin{figure}[htbp]
\centering
\includegraphics[width=9cm]{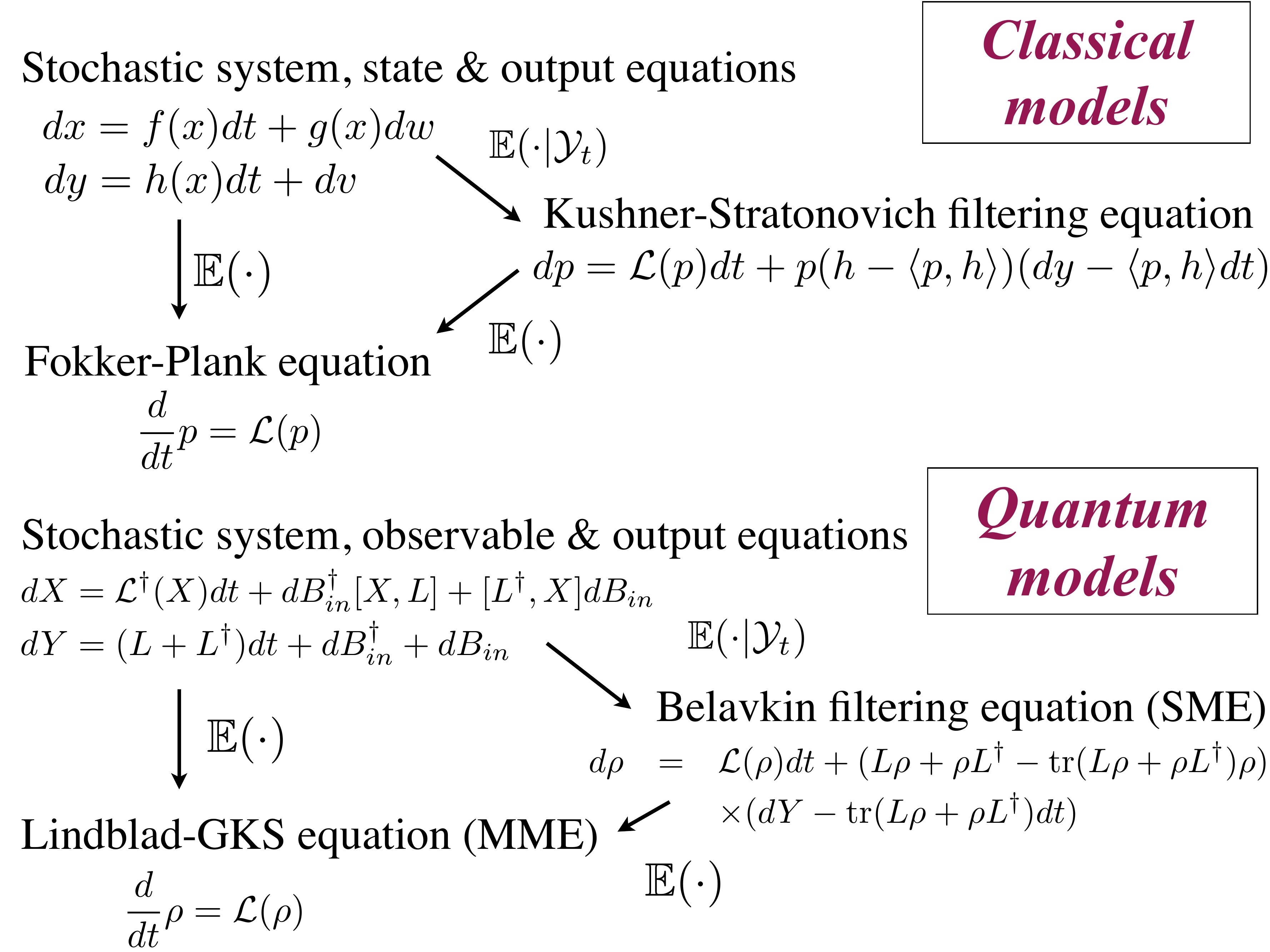} 
\caption{A parallel between the models of classical and quantum stochastic dynamics, and the role of expectations. Adapted from \cite{JamesLN}.}
\label{classical-quantum} 
\end{figure}

\begin{remark} [Classical and quantum correspondences] Notice how \eqref{eq:SME} is equivalent to a semigroup generator in Lindblad form plus a nonlinear, stochastic term accounting for the state update due to the innovation in the measurement record. It plays the role, in this non-commutative context, of the classical Kushner-Stratonovich equation \cite{Doherty2000}. Because the stochastic term is a martingale, the average over the noise trajectories leaves us with a MME in the form \eqref{eq:master2}, which plays the role of the classical Fokker-Plank equation. Of course, one can derive the generator by directly taking the expectation over the noise trajectory on the state equation, which makes sense in particular when no observation on the system is available. These correspondences are depicted in Fig.~\ref{classical-quantum}, with reference to the quantum models emerging from the homodyne-detection scheme. 
In fact, in the classical setting the filtering equation for a stochastic system is derived by conditioning on the past history of its output (represented by ${\cal Y}_t$, which is a $\sigma$-algebra in the classical case and an equivalent commutative von Neumann algebra in the quantum case) and the associated semigroup generator by averaging over the Martingale part.
\end{remark}

\section{Controllability of Quantum Systems}\label{sec:controllability}
Control systems such as the Schr{\"o}dinger equation or the MME, driven by unitary controls, are bilinear control systems.
A wealth of results is available to characterize the controllability properties of these systems, see {\em e.g.} \cite{Jurdjevic1} for a survey and \cite{Huang1983} for an early application to quantum control.
In this Section these results are applied to both closed and open quantum systems.

\subsection{Controllability of the Schr{\"o}dinger equation}
For a system like \eqref{eq:Schrod2}, the {\em reachable set} from $ \ket{\psi_0}$ at time $ t$ is defined as 
\[\begin{split}
{\cal R}(\ket{\psi_0}, t) & = \Big\{ \ket{\psi(t)} \in \mathbb{S}^{2N-1} |\; \exists\; u(\cdot ) \text{ s.t. } \\ & \ket{\psi(t)}   = {\cal T } {\rm exp }\left( \int_0^t -i (H_0 + u(s) H_1 ) ds \right) \ket{\psi_0 }  \Big\}.
\end{split}\]
In an analogous fashion, $ {\cal R}(\ket{\psi_0}, \leqslant T) = \bigcup_{0\leqslant t \leqslant T } {\cal R}(\ket{\psi_0}, t) $ and $  {\cal R}(\ket{\psi_0} ) =  \bigcup_{0\leqslant t \leqslant \infty } {\cal R}(\ket{\psi_0}, t) $.

The system \eqref{eq:Schrod2} is called (globally) controllable at $ \ket{\psi_0 } $ if $  {\cal R}(\ket{\psi_0} ) = \mathbb{S}^{2N-1},$ and Small-Time Locally Controllable (STLC) if $  \ket{\psi_0 }\in {\rm int} (  {\cal R}(\ket{\psi_0}, \leqslant t )) $ $ \forall\, t>0 $. Analogous definitions hold for the unitary propagator $ U$. Denoting $ {\cal R}_U (\cdot ) $ the corresponding reachable sets, the property of right invariance of \eqref{eq:unitary_propag1} implies invariance of the group-evaluated reachable sets: $ {\cal R}_U ( U_0 ) =  {\cal R}_U (I ) U_0 $ $ \forall\, U_0 \in SU(N)$. 
A remarkable feature of compact Lie algebras is that the controllability properties admit an infinitesimal characterization.
For $SU(N)$ in particular we have the following theorem, originally due to \cite{Jurdjevic3}.
Denote $ {\rm Lie} \{ A,\, B \} $ the Lie algebra generated by the matrices $A$ and $ B$.
\begin{theorem} 
\label{thm:jurdj-susss} The system \eqref{eq:unitary_propag1} is controllable {\em iff} $ {\rm Lie} \{ -i H_0 , \, -i H_1 \} = \mathfrak{su}(N)$. 
\end{theorem}

For general nonlinear systems, the Lie Algebraic Rank Condition (LARC) of Theorem~\ref{thm:jurdj-susss} is characterizing only the accessibility property, i.e., whether the smallest Lie algebra spanned by the vector fields of the system coincides (locally) with the entire tangent space of the system.
For a Lie group like $ SU(N)$, however, due to its compactness, the free Hamiltonian $ -i H_0 $ is not an ``obstacle'' to controllability for sufficiently long times, although a system like \eqref{eq:unitary_propag1} fails to be STLC.
This result provides a sufficient condition for the controllability of the Schr{\"o}dinger equation.
In fact, since the action of $ SU(N)$ on $ \mathbb{S}^{2N-1} $ is transitive, we know from the theory of bilinear control systems \cite{Boothby3,Jurdjevic1} that the controllability of the system lifted to the Lie group \eqref{eq:unitary_propag1} implies the controllability of \eqref{eq:Schrod2} on the corresponding homogeneous space.
The condition is sufficient (``operator controllability'' implies ``pure state controllability'' in the language of \cite{Albertini1}) but not necessary. 
In fact, there exist cases in which $ {\rm Lie}\{  -i H_0, \, -i H_1 \} \subsetneq \mathfrak{su}(N)$ but the corresponding reachable set $ {\cal R}_U $ still acts transitively on the sphere $ \mathbb{S}^{2N-1}$, see \cite{Albertini1} for details.
Owing to the semisimple nature of the Lie algebras involved, these cases are however {\em rare}. 
In fact, in semisimple Lie algebras we have the following generic result.
\begin{theorem} 
\label{thm:kuranishi}
(\cite{kuranishi1}) The set of pairs $ A, \, B \in \mathfrak{su}(N) $ for which $ {\rm Lie} \{ A, \, B \}  = \mathfrak{su}(N) $ is open and dense in $ \mathfrak{su}(N) $.
\end{theorem}
Putting together Theorems \ref{thm:jurdj-susss} and \ref{thm:kuranishi}, we have genericity of controllability on $ SU(N)$ and hence on $ \mathbb{S}^{2N-1}$ due to transitivity \cite{Jurdjevic2}.
\begin{corollary}
\label{cor:almost_always}
The system \eqref{eq:Schrod2} is controllable for almost all pairs $ H_0 $ and $ H_1 $. 
\end{corollary}
As was first observed in \cite{Ramakrishna4}, Corollary~\ref{cor:almost_always} is the infinitesimal counterpart of the following quantum information result, valid for (nonlocal) quantum ``gates'', i.e., discrete-time unitary operators such as $ U_0= {\rm exp} ( -i H_0 \tau )  $ and $ U_1 = {\rm exp} ( -i (H_0 + u_1 H_1 ) \tau )  $ for some given $ \tau $ and $ u_1 $.
\begin{theorem}
\label{thm:lloyd}
(\cite{Lloyd1}) Almost any pair of unitary gates $ U_0 $ and $ U_1 $ is universal.
\end{theorem}
In the quantum information literature ``universal'' is used as a synonymous of operator controllability.
Thm.~\ref{thm:lloyd} says that by composing arbitrarily many times the unitaries  $ U_0 $ and $ U_1 $ it is almost always possible to reach any point of $ SU(N)$, i.e., $ {\cal R}_U (I) = SU(N)$. 
Controllability of $ \ket{\psi} $ then follows from transitivity of $SU(N)$ on $ \mathbb{S}^{2N-1}$.

\begin{example} [Two-level system]
For $ \ket{\psi} \in \Hi^2 ,$ in the energy eigenvector basis, up to a scalar, the free Hamiltonian is of the form $ H_0 = \Delta \sigma_z ,$ with $\Delta$ a scalar parameter. A typical control Hamiltonian is $ H_1 = \sigma_x $,
which enables transitions between the two energy levels. Controllability is then ensured, since the first commutator yields 
$[-i\Delta\sigma_z,-i\sigma_x]=-i2 \Delta \sigma_y$,
and $\{-i\sigma_{x,y,z} \}$ generate the full Lie algebra $\mathfrak{su}(2).$
\end{example}

A particular feature of a semisimple Lie algebra such as $ \mathfrak{su}(N) $ is that the Lie brackets never vanish as we increase the level of bracketing. 
Hence the verification of Theorem~\ref{thm:jurdj-susss} by brute force calculations becomes rapidly cumbersome. 
Alternative, easy-to-check sufficient conditions for controllability can be given for particular families of generators \cite{Turinici1,Cla-contr-root1}.
Consider again the basis in which $ H_0 $ diagonal: in this basis, in fact, the control Hamiltonian $ H_1 $ contains the couplings between the energy levels $ \eneig_j $ of the system. 
The {\em graph} ${\cal G}_{H_1} $ associated to $ H_1 = [
b_{ij} ] $, is the pair ${\cal G}_{H_1} = ( {\cal N}_{H_1}, \, {\cal A}_{H_1} )$
where $ {\cal N}_{H_1}= \left\{ 1, \ldots , n \right\}  $ are the nodes and $ {\cal A}_{H_1} = \left\{ (i, \, j ) \; | \; b_{ij} \neq 0 \right\}$ the directed edges connecting the nodes.
$ {\cal G}_{H_1} $ is said (strongly) {\em connected} if for all pairs of nodes in $ {\cal N}_ {H_1}$ there exists a (directed) path in ${\cal A}_{H_1} $ connecting them.
If $H_0$ is diagonal, a necessary condition for controllability of \eqref{eq:Schrod2} is that $ {\cal G}_{H_1} $ be connected.
In fact, if $ {\cal G}_{H_1} $ is not connected, there exist nontrivial invariant
subspaces of $\mathfrak{su}(N) $ that are simultaneously $H_0$-invariant
and $H_1$-invariant. Thus the system cannot be controllable.
Physically, the connectivity of $ {\cal G}_{H_1} $ means that it is in principle possible to transfer population from any energy level $ \eneig_j $ to any other  $ \eneig_k $.
When the system has degenerate transitions (i.e., it is not strongly regular), however, not all population transfers may be possible, because different transitions may be excited by the same control field, meaning that controllability may be lost. 
The following Theorem, inspired by the literature on geometric control of systems on semisimple Lie groups \cite{SilvaLeite2}, provides sufficient conditions for controllability.
\begin{theorem}
\label{thm:suff-contr-Schr1}
(\cite{Turinici1,Cla-contr-root1}) 
If $ H_0 $ is strongly regular and $  {\cal G}_{H_1} $ is connected then the system \eqref{eq:Schrod2} is controllable.
\end{theorem}
Since both strong regularity of $ H_0 $ and connectivity of $ {\cal G}_{H_1} $ are valid in an open and dense set of generators of $ \mathfrak{su}(N)$, Theorem~\ref{thm:suff-contr-Schr1} is generically verified.
The condition of Theorem~\ref{thm:suff-contr-Schr1} is not necessary and can be weakened in several ways \cite{Cla-contr-root1}.
For example, one can ask for the Bohr frequencies to be all different only on the graph of $ H_1 $, or, alternatively, that the graph of $ H_1 $ remains connected when restricted to the set of arcs $ (j, k ) $ that correspond to all different Bohr frequencies $ \eneig_j - \eneig_k $.

The extension of these results to the Liouville - von Neumann equation is straightforward.
In \eqref{eq:liouville2}, if $ \rho(0) \in \mathcal{C}$ then  $ \rho(t) \in \mathcal{C}$ $ \forall \, t\geqslant 0$, and the reachable set of the bilinear control system is at most $ \mathcal{C}$.
Since $ \mathcal{C} $ are homogeneous spaces of $ SU(N)$, controllability results analogous to those for the Schr\"odinger's equation hold.

\subsection{Controllability of the MME}
\label{subs:control-MME}
The main feature of the MME is to capture all the possible time-invariant infinitesimal generators that leave $ {\mathfrak D}(\Hi)$ invariant for the evolution. Under proper assumptions on the interaction with the environment \cite{alicki-lendi}, one can add Hamiltonian controls to \eqref{eq:master2}, obtaining
\begin{equation}\label{eq:mastercont}
\ddt \rho(t)=-{i}[H,\rho(t)]-i\sum_k u_k[H_k,\rho(t)] +\sum_{k} {\cal D}(L_k,\rho(t)).
\end{equation}
The dissipation term $ {\cal L}_D=\sum_{k} {\cal D}(L_k,\rho(t))$ is not unitary, hence the evolution of the MME \eqref{eq:mastercont} is no longer invariant on a complex flag manifold $ \mathcal{C} $ and no longer isospectral. 
Correspondingly, the Lie (semi)group acting on $ \rho$ for the MME must be larger than $ SU(N)$. 
In particular, when $ {\cal L}_D$ is not unital the integral group in which the QDS \eqref{eq:mastercont} is ``sitting'' must act in an affine way on $ \rho$.
Infinitesimally, affine actions are more easily characterized using vectors of coherences and adjoint representations, as in \eqref{eq:lindblad-bloch}. 
A Lie algebra $ \mathfrak{g}$ acting affinely on a vector space is said to have a semidirect sum structure, $ \mathfrak{g} =   \mathfrak{g}_1 \circledS V$, where  $ \mathfrak{g}_1$ is a linear (homogeneous) Lie algebra and $ V$ a vector space \cite{Marsden2}. 
When instead $ {\cal L}_D$ is unital, then the non-homogeneous part of $ \mathfrak{g}$ is absent: $ \mathfrak{g} =   \mathfrak{g}_1 $.
Computing accessibility for an $N$-level system, for example, is a matter of verifying which of the Lie algebras $ \mathfrak{g}$ are acting transitively on $ \mathbb{R}^{N^2 -1 }$ ($ N^2-1$ is the dimension of the coherence vector), see \cite{Cla-contr-open1,Kurniawan07Dynamics}.
We report here the main result of \cite{Cla-contr-open1}, where an analysis of controllability is developed in detail. 
\begin{theorem}
\label{thm:acc-contr-N}
The system \eqref{eq:mastercont} is
\begin{enumerate}
\item accessible for $ \mathfrak{g} = \mathfrak{gl} ( n , \mathbb{R} )$ or $ \mathfrak{g} = \mathfrak{gl} ( n , \mathbb{R} ) \circledS \mathbb{R}^n $;
\item never STLC;
\item never controllable in finite time;
\item never controllable in $ \mathfrak{D}(\Hi) $ for $ {\cal L}_D $ unital.
\end{enumerate}
\end{theorem} 
The lack of small-time controllability follows from the fact that the flow of \eqref{eq:mastercont} contains directions, infinitesimally described by $ {\cal L}_D $, which are not unitary, hence cannot be reversed by means of unitary controls.
Therefore, even under arbitrary unitary control, the infinitesimal object generated by the vector fields is only a Lie wedge (roughly speaking, a non-pointed cone, see \cite{Dirr09Lie}), and the corresponding reachable sets $ {\cal R}(\rho(0), \leq t) $ form only a Lie semigroup.
While STLC is missing for any $ {\cal L}_D $, the two cases of unital and ``affine'' $ {\cal L}_D $ are characterized by different properties for long times. 
In particular, for $ {\cal L}_D $ unital the reachable sets and the eigenvalues of $ \rho$ can be completely characterized in terms of monotonicity properties \cite{Cla-contr-open1,Yuan10Characterization}.
If $ \mu_{i,j}$, $ i=1,2$, $j=1, \ldots N$ are the eigenvalues of $ \rho(t_i)$ sorted in decreasing order, then let the symbol ``$\prec$'' denote majorization: $ \sum_{j=1}^k \mu_{2,j} \leq \sum_{j=1}^k \mu_{1,j}$, $ k = 1 , \ldots, N-1 $, with equality for $ k=N$, see \cite{Yuan10Characterization}.
\begin{proposition}
\label{prop:reach-unital}
If the system \eqref{eq:mastercont} is accessible and if $ {\cal L}_D$ is unital, then 
\begin{enumerate}
\item $ {\cal R}(\rho (0) ,  \, \leq t_2 ) \supsetneq {\cal R}(\rho (0),  \, \leq t_1 ) $ $\quad  \forall \,t_2 > t_1 > 0$;
\item $ \Phi(\rho(t_2)) \prec  \Phi(\rho(t_1)) $ $\quad \forall \, t_2 > t_1$.
\end{enumerate}
\end{proposition}
Together Theorem~\ref{unique-attractive} and Proposition~\ref{prop:reach-unital} provide a complete characterization of the contraction behavior of $ \rho$ towards the steady state $ \rho = I/N $ when this is unique and $ {\cal L}_D $ is unital.

\begin{example}[Two-level system with $ {\cal L}_D $ unital]
For any $ H_0 $ and $ H_1 $ such that $ {\rm Lie}( -i H_0, \, -i H_1 ) = \mathfrak{su}(2)$, and $ \Gamma$ such that $ \Gamma $ invertible, $ \Gamma $ not proportional to $I$, $ \tr{\Gamma} \neq 0$, the undriven system has $ \bm{\rho} = 0$ as a globally asymptotically stable attractor. 
The Lie algebra is $ \mathfrak{g} = {\rm Lie} ( B_{H_0} + \Gamma, \, B_{H_1} ) = \mathfrak{gl}(3, \mathbb{R}) $ and the system is accessible.
The reachable sets are $ {\cal R}(\bm{\rho} (0)) = \{ \bm{\rho} \in \mathbb{R}^3  \text{ s.t. } \| \bm{\rho} \| \leqslant \| \bm{\rho} (0) \| \} $, i.e., the solid balls of radius equal to the initial condition.
In this case the purity $  \| \bm{\rho}  \|^2 $ is a quadratic Lyapunov function centered in the completely random state also for the controlled system.
\end{example}

\begin{example} [Two-level atom with decay]
As the unital part of $ {\cal L}_D $ (i.e., $\Gamma$ in \eqref{eq:example-affine}) is invertible, the system \eqref{eq:sp-em} has a unique fixed point and it is a globally asymptotically stable equilibrium for the uncontrolled MME, see Theorem~\ref{unique-attractive}. 
The fixed point is $ \rho_f = \left( B_{H} + \Gamma \right) ^{-1} g $, which is the pure state $\rho_f=\ket{1}\bra{1}$ in absence of controls.
Constant non-zero coherent controls can be used to modify $ \rho_f,$ but the equilibrium becomes a mixed state.
A global order in the reachable sets, similar to Proposition \ref{prop:reach-unital}, cannot in general be established, see \cite{Cla-contr-open1,Cla-contr-open4} for examples. 
\end{example}

\section{Synthesis methods for open-loop control}\label{sec:openloop}

For unitary controls, the spectrum of synthesis techniques is overwhelmingly wide, and different techniques are better suited for different experimental systems. It is well beyond the scopes of this tutorial to enter the details of each of them, so we are constrained to a brief survey of the most notable categories of control design strategies.

It is also worth remarking that, with the Hamiltonian control methods discussed here, the presence of an (uncontrollable) environment usually poses significant limits to the achievable tasks, as the controllability analysis of Section~\ref{subs:control-MME} revealed. However, the capability of directly engineering {\em open-system} dynamics can drastically change the scenario, and significantly extend the set of addressable control problems. For example, it is clear from Section~\ref{subsec:stability} that these techniques can in principle allow for asymptotic state stabilization without the necessity of resorting to feedback methods. 
The scope of the so-called {\em environment engineering} methods has been investigated under a number of different assumptions, see {\em {\em e.g.}} \cite{viola-engineering,poyatos,davidovich,Verstraete2009,schirmer-markovian,ticozzi-generic}. 

\subsection{Optimal control algorithms}\label{optimal}
As already mentioned in the Introduction, optimal control methods are widely used in laser-driven molecular reactions \cite{Peirce1988,Dahleh1}. 
A recent control-oriented introduction to quantum optimal control can be found in \cite{dalessandro-book}. 
A cost function often used in this context is for example
\[
J = \langle \psi(T)  | M | \psi(T)  \rangle - \int_0^T u^2 d \tau 
\]
i.e., simultaneously maximize the expectation value of a target observable $M$ and minimize the ``fluence'' (energy) of the control. 
A necessary condition for optimality is that the first variation of $ J $ subject to the Schr\"odinger equation \eqref{eq:Schrod2} vanishes.
The corresponding Euler-Lagrange equations are then given by \eqref{eq:Schrod2}, together with the adjoint equation for the co-state $ \chi $
\[
\begin{cases}
\dot{\chi} & = -i (H_0 + u H_1 )  \chi  \\
\chi(T) & = M  \ket{\psi(T)}
\end{cases}
\]
and the control $
u (t) = - {\rm Im}\left[  \langle \chi(t)  | H_1 | \psi(t)  \rangle \right]$,
see {\em e.g.} \cite{Zhu1998385} for the details.
These Euler-Lagrange equations are normally solved numerically, and a plethora of numerical schemes have been developed for this and similar formulations, which often enjoy monotonically convergent properties \cite{Zhu1998385,Maday20038191}.
In particular, iterative algorithms for generating piecewise constant controls can be subdivided into Korotov-type (in which each time slice is updated sequentially) \cite{Krotov83Iteration} and GRAPE-type (in which all time intervals are updated concurrently) \cite{Khaneja05optimal}.
These methods are surveyed and compared in \cite{Machnes10Comparing}. 
Alongside the numerical schemes, a few analytical results have appeared in recent years, based on the explicit calculation of the extremal solutions of the Pontryagin Maximum principle \cite{Boscain1}, applicable only to low-dimensional systems.

\subsection{Lyapunov-based design}
An alternative approach for the generation of open-loop trajectories consists in using a ``feedback on the model'' strategy, i.e., on solving the control problem as a feedback problem and using the trajectories produced by the simulator as open-loop controls. 
This allows the complete bypassing of the issues related to the complexity of the optimal control schemes. 
For state vectors, this approach is followed for example in  \cite{Ferrante1,Mirrahimi2} and for density operators in \cite{Cla-qu-ens-feeb1,schirmer-wang-2010a}.
For a closed system such as \eqref{eq:liouville2}, with $ H(t,u) = H_0 + u H_1 $, the Lyapunov design relies essentially on a Jurdjevic-Quinn condition and LaSalle invariance principle (see \cite{Cla-qu-ens-feeb1}), i.e., on choosing as Lyapunov function a distance from a target state $ \rho_d $ ($ \rho, \, \rho_d \in \mathcal{C}$, i.e., isospectral)
\beq
V ( \rho_d, \rho) = \frac{1}{2} \| \rho_d - \rho \| ^2 = \um\tr{\rho_d ^2} - \tr{\rho_d \rho} +\um\tr{\rho ^2}.
\label{eq:Lyapunov}
\eeq
Assuming for simplicity that $ \rho_d $ is a fixed point of the free Hamiltonian, $ [ -i H_0, \, \rho_d ]=0 $, noting that the purity is an invariant for unitary evolution and differentiating \eqref{eq:Lyapunov} one has 
\beq
\dot V = u \tr{ [ -i H_1 , \, \rho_d ] \rho } ,
\label{eq:derLyapunov}
\eeq
which is readily made into a negative semidefinite function by choosing 
\beq
u = - \tr{ [ -i H_1 , \, \rho_d ] \rho } .
\label{eq:controlLyapunov}
\eeq
Local asymptotic stability can be achieved when the following Kalman-like rank condition is satisfied by the system linearized around $ \rho_d$ \cite{Cla-qu-ens-feeb1,Mirrahimi2}
\[\dim\Big( {\rm span} \Big(\ad_{-i H_0 }^\ell [-i H_1, \rho_d ], \, \ell =0, 1, \ldots  \Big) \Big) = \dim \left( T_{\rho_d}\mathcal{C} \right),
\]
where $ \ad_A B = [ A, \, B] $, and $ T_{\rho_d } \mathcal{C} $ is the tangent space at $ \rho_d $. 
For strongly regular $ H_0 $ and $ {\cal G}_{H_1} $ fully connected this condition is always verified. 
The attractivity achieved by this feedback design can never be global for topological reasons: $ \mathcal{C} $ is a compact manifold without boundary, hence it is not continuously contractible to a point \cite{Bhat1}.
The obstructions to global stabilizability are spurious equilibria $ \rho_p \in \mathcal{C}$ corresponding to ``antipodal'' points of $ \rho_d $ in $\mathcal{C}$: in a basis in which $ \rho_d $ is diagonal, $ \rho_p $ are permutations of the diagonal entries of $ \rho_d $ (for $N=2$, $ \rho_p $ is indeed the antipodal point to $ \rho_d $ on $\mathbb{S}^2 $). 
In particular, for ``generic'' $\mathcal{C}$ (characterized by a spectrum $ \mu_1, \ldots, \mu_N$, $ \mu_j\neq \mu_k $), $V$ is a Morse function \cite{schirmer-wang-2010a}, i.e., $ \dot{V}(\rho_d, \rho ) \neq 0 $ except for the $ N! $ hyperbolic critical points of $V$.
Of these, one ($ \rho_d $) is a global minimum (and a sink for the closed-loop dynamics \eqref{eq:liouville2}-\eqref{eq:controlLyapunov}), another is a global maximum (a source), and the remaining are all saddle points. 
All closed loop trajectories converge to $ \rho_d $ except for a zero-measure set (stable submanifolds of the saddle points) \cite{schirmer-wang-2010a}.

\subsection{Average Hamiltonian Methods and Dynamical Decoupling}\label{decoupling}
A large family of pulsed-control protocols that rely on average Hamiltonian techniques have been developed for the control of closed and open quantum systems. They are inspired by pulse sequences in NMR \cite{Mehring1}, and have been successfully applied to other settings (from solid-state \cite{petta-science-mod} to optical systems \cite{vitali-flying}) for suppression of undesired noise components and design of quantum dynamics on the systems of interest \cite{viola-gates}. We here only briefly sketch the ideas underlying the basic {\em Dynamical Decoupling} (DD) technique for suppressing the interaction with the environment, referring the reader to {\em e.g.} \cite{viola-dd,ticozzi-feedbackDD,viola-gates} for more details and recent developments.

Consider the bipartite system/environment setting described in Section \ref{subsec:opensys}, with the Hamiltonian described by \eqref{hamiltonianopen}.
We assume the interaction Hamiltonian $H_{\Si\Ei}$ to be of the form $H_{\Si\Ei}=\sum_{k=1}^{q}S_k\otimes E_k,$ with $S_k$ being operators acting on the system space $\Hi_\Si$ and $E_k$ being an operator acting on $\Hi_\Ei$.
To control the evolution, we rely on a time-dependent control
Hamiltonian ${H}_c(t)$ that acts on the system alone. In most DD protocols,
$H_c(t)$ is taken to be cyclic, namely, the associated propagator $U_c(t)={\cal T}e^{-i \int_0^t H_c(s)ds}$ is
periodic:

\begin{equation}U_c(t)=U_c(t+T_c),\end{equation}

\noindent for some $T_c>0$. This implies that
${U}_c(nT_c)=I,\, n\in\N$.
Suppose we choose a piecewise constant control
propagator: ideally, it can be generated by a
discrete-time sequence of {\em impulsive, unbounded controls}. Let us set $U_c(t)\equiv G_j,\quad j\Delta t\leq t\leq
(j+1)\Delta t,$ with
$\mathfrak{G}=\{G_j\}_{j=1}^{n_g}$ a finite set of unitary
operators and $\Delta t:=T_c/n_g$. 
With these definitions, the {\em full} propagator ${U}(t),$ accounting for both the free Hamiltonian $H_{tot}$ and the control action $H_c(t)$, can be expanded
in Magnus series \cite{magnus}, and for $t=T_c$ one gets:
\begin{equation}\label{mag}
{U}(T_c):=e^{-i(\bar H^{(0)}+\bar H^{(1)}+\bar
H^{(2)}+...)T_c},
\end{equation}
where the first-order term is the average Hamiltonian:
\begin{eqnarray}\bar{H}^{(0)}&=&\frac{1}{T_c}\int_0^{T_c}\left(\sum_{k=1}^qU^\dag_c(t)S_kU_c(t)\otimes
E_k\right)ds\nonumber\\\label{Hmean}
&=&\sum_{k=1}^q\left(\frac{1}{n_g}\sum_{j=1}^{n_g}G_j^\dag S_k
G_j\right)\otimes E_k.
\end{eqnarray}
The evolution is then evaluated after $K$ control cycles are applied in a fixed interval $[0,T]$, $T=KT_c.$ Considering the
(ideal) limit $K\rightarrow\infty$ with fixed $T$, it can be shown under fairly general conditions that the higher order terms in (\ref{mag}) become negligible, and the full propagator can be approximated by the {\em average Hamiltonian} ${U}(T)\approx e^{-i\bar H^{(0)}T}.$

In the early formulation of DD \cite{viola-dd},  $\mathfrak{G}$ is chosen to be a
(finite) group and it can be shown that the expression between
brackets in \eqref{Hmean} defines the projection of every
$S$ onto the \emph{commutant} of $\mathfrak{G}$. Properly choosing
$\mathfrak{G}$ one can then cancel the
effect of the average Hamiltonian $\bar{H}^{(0)}$. 
Hence, this control strategy has been extensively used to make the interactions with the environment negligible for the system in the fast
control approximation.
Building on these basic ideas, a wide spectrum of decoupling-like strategies have been devised, including effective cycles using bounded control pulses, concatenated protocols to address higher order corrections, optimization of time intervals, and techniques for engineering the desired dynamics while decoupling the system from the environment \cite{viola-gates}.

\subsection{Learning algorithms for control design}

As we recalled in Section \ref{introduction}, the control of chemical reactions has been one of the first tasks of interests for quantum control: learning algorithms have been successfully used to iteratively design a laser field in order to obtain a desired population transfer, to manipulate fluorescence signals and photodissociation products and other applications \cite{Rabitz1,expcont2-mod,expcont3}, see \cite{Brif2010Control} for a recent survey. While on each cycle the control is in fact in open-loop, a closed-loop iterative method is employed to generate optimal control actions. Starting with a trial control input (may be as well a quasi-random field \cite{Rabitz1}), the typical iteration entails: (i) a fast experimental test of the last generated field, with in particular a measurement or evaluation of some cost functional; and (ii) an algorithm that based on the performance of the previous control choices provides the new trial field. 
In \cite{Rabitz1}, in particular, genetic algorithms have been employed to seek an optimal control field, with the cost functional depending only on the final state $\rho(T),$ namely $J(T)=\tr{(\rho_d-\rho(T))^2}.$ Other more recent approaches commonly employ steepest descent and other optimal control algorithms, see Section \ref{optimal}.

As a result, one finds that it is somehow surprisingly easy to find effective control fields, at least when the control constraints are not strict. This and other considerations have led to an extensive analysis of the ``quantum control landscapes'', that is, of the topology of the surfaces defined by the cost functional when the control parameters are varying on the available set (see {\em e.g.} \cite{landscapes2}). In the absence of control constraints they are conveniently ``trap-free'', namely steepest descent algorithms lead to global optima. Under practical restrictions, however, the situation may become more complicated.

\section{Synthesis Methods for Feedback Control}\label{sec:closedloop}

Feedback control methods for quantum systems can be grouped in two main families. The first one, that includes {\em measurement-based feedback} methods, involves some kind of measurement on the system, direct or indirect, that produces a classical output signal, which is then fed to a (classical) controller in order to determine the control input. 

The second, on the contrary, does not require the transition to classical information: typically, the system of interest is dynamically entangled, and hence shares its quantum information, with another quantum system, e.g. a quantum field acting as an {\em output signal}, which in turn is fed to a fully quantum network, acting as a {\em quantum controller}. The latter is engineered to feed back an input field to the system in order to achieve the desired control task. The overall evolution of the system-signals-controller network being unitary, this approach is typically referred as {\em coherent quantum feedback}. The first example of a coherent quantum feedback is the all-optical feedback presented in \cite{wiseman-alloptical}. Since then, a considerable control-oriented literature has been developed, see e.g.
\cite{transferfunction-1,transferfunction-2,james-2008,gough-product,james-passivity,nurdin-coherent}.
In the following, we present the most common approaches to measurement-based feedback control, and briefly introduce the reader to the language of the relatively unexplored world of quantum feedback networks.

\subsection{Output feedback and feedback master equation}

Let us consider an evolution of the form \eqref{eq:SME}. As for a classical SISO control system, the simplest form of feedback control is indeed {\em output feedback}, where the control input is a static function of the output signal. If we assume that there is no delay due to the feedback loop, the resulting evolution is Markovian. Output feedback techniques have a wide range of applications, including protecting entanglement in optical cavities \cite{HK97}, spin-squeezing \cite{TMW02}, cooling of trapped ions \cite{ions2-mod}, engineering quantum information \cite{ticozzi-QDS}, error correction \cite{ahn-feedback}, and line-width narrowing of atom lasers \cite{WT01}. 

Consider, as the feedback action, the Hamiltonian $H_f(t)=FdY_t,$ where $dY_t$ is the output signal and $F=F^\dag$ a time-invariant Hamiltonian. Here we assume perfect detection efficiency $\eta=1,$ but the model can be extended to the non-ideal case \cite{Barc09}. Computing the infinitesimal evolution resulting from the measurement action (instantaneously) followed by the feedback action 
leads to the Wiseman-Milburn {\em Markovian Feedback Master equation}
(FME) \cite{wiseman-milburn,wiseman-feedback}: \beq\label{eq:MME} \ddt
\rho_t=-i[H_{tot},\rho_t]+{\cal
D}(L - iF,\rho_t),\eeq
\noindent
where the new Hamiltonian is $H_{tot}=H+ \frac{1}{2}(FL+L^\dag F).$
Note that the resulting equation is indeed in Lindblad from, and it thus represents a valid generator for a quantum Markovian semigroup, that can be thought of as the average closed-loop dynamics with respect to the noise trajectories. The dynamics we obtain are still in the form of a QDS generator, but the noise operator depends on {\em both the measurement and control} actions.

The primary task of feedback control, in the classical as well as in the quantum domains, is to attain stabilization of a desired state. The problem for Markovian dynamics has been extensively studied, with a focus on a single two-level system under different control assumptions \cite{wang-wiseman,wiseman-book,Barc09,ticozzi-markovian,ticozzi-QDS}. 
A general pure-state stabilization problem for controlled Markovian dynamics described by FMEs is addressed by the following result, which is a specialization of the more general subspace-stabilization problem of \cite{ticozzi-markovian}. 

\begin{theorem} \label{purestateGAS}
For any measurement operator $L$, there exist a feedback
Hamiltonian $F$ and a Hamiltonian compensation $H_c$ able to make
an arbitrary desired pure state $\rho_d$ GAS for the FME \eqref{eq:MME}
{\em iff}: 
\beq\label{stabilize2}[\rho_d,(L+L^\dag)]
\neq0.\eeq
\end{theorem} 

The proof of Theorem \ref{purestateGAS} yields a constructive
algorithm for designing the feedback and correction Hamiltonians
needed for the stabilization task.

\begin{example}[FME state-preparation for two-level systems]
Consider a two-level FME of the form \eqref{eq:MME}. Assume $H=\sigma_z$, and $L=\frac{1}{2}\sigma_x$. Without loss of generality, let the target
state be written as $\rho_d=\diag(1,0)$, and write, with respect to the same basis,
$$ L-iF=\hat{L}=\left(%
\begin{array}{cc}
\hat{l}_{S} & \hat{l}_{P} \\
\hat{l}_{Q} & \hat{l}_{R} \\
\end{array}%
\right),\quad H_{tot}=\left(%
\begin{array}{cc}
h_{S} & h_{P} \\
h_{P}^* & h_{R} \\
\end{array}%
\right). $$
The pure state $\rho_d=\diag(1,0)$ is a globally 
attractive, invariant state for a two-dimensional quantum system
{\em iff}:
\beq i h_{P}-\frac{1}{2} \hat{l}_{S}^*\hat{l}_{P}=0,\label{eq:c1}\quad \hat{l}_{Q}=0, \quad \hat{l}_{P}\neq 0.\eeq
It is easy to show that engineering $\hat{L}=\sigma_+$  indeed stabilizes $\rho_d$. This can be attained by properly choosing the feedback Hamiltonian.
The measurement operator $L$ satisfies the condition for stabilizability of Theorem \ref{purestateGAS}, hence $\rho_d$ is stabilizable. 
It is easy to see that the needed feedback Hamiltonian is $F=-\frac{1}{2}\sigma_y$. In this case $\frac{1}{2}(FL+L^\dag F)=0$, and no additional Hamiltonian correction is required, thus $H_{tot}=H$.  
\end{example}

\subsection{Filtering-based feedback}

A richer control strategy with respect to the simple output feedback can be devised by exploiting the fact that the SME \eqref{eq:SME} describes the dynamics for the state estimate, conditioned on the measurement record and depending on the control law. Assume that it is possible to integrate \eqref{eq:SME} in real time, starting from some {\em a priori} initial condition: at each $t$ an updated estimate of the system state $\rho_t$ is then available. As in the classical case, this can be used to design a state-feedback law $f(\rho_t)$. The generic stability of the filter has been discussed in \cite{vanhandel-phd}, and a separation principle has been shown to hold \cite{vanhandel-separation}. The system dynamics, assuming that no dissipative component other than the  measurement is present, is typically assumed to be of the form:
\begin{equation}
\begin{split}
d\rho_t=&\big( -i [ H_0+f(\rho_t)H_1,\rho_t] 
+{\cal D}(L,\rho_t)\big) dt\\& + \sqrt{\eta}\,{\cal G}(L,\rho_t) d W_t, \\
\end{split}
\label{eq:SMEcont}
\end{equation}
with fixed Hamiltonians $H_0$ and $H_1,$ and $0<\eta\leq 1$ accounting for limited efficiency in the homodyne-type detection.
Due to the topological structure of the state space, the feedback design generally presents significant difficulties, and cannot be pursued by standard methods \cite{Florchinger1994,Florchinger1995}. 
A first way around these difficulties is offered by numerical optimization, in particular semi-definite programming \cite{vanhandel-feedback}. Consider the same two-level system as in \eqref{SME3}, with Hamiltonian $H(t)=  u(t)\sigma_{y}$ (for simplicity we assume $\Delta=0$), where the control $u(t)$ is the amplitude of the magnetic field in the $y$-direction. That is, we have:
\begin{equation}\label{SME5}\begin{split}
d\rho_t=&-i [u(t)\sigma_y, \rho_t]dt+\nu(\sigma_{z}\rho_t\sigma_z- 
\rho_t)dt\\&+\sqrt{\eta\nu}(\sigma_{z}\rho_t+\rho_t \sigma_{z}-2\tr{
\sigma_{z}\rho_t)\rho_t}dW_t,
\end{split}
\end{equation}
or, in terms of the Bloch vector:
\begin{equation}
\label{eq:bloch}
\begin{split}
	dx_t &= (u(t)z_t-\tfrac{1}{2}\nu x_t)\,dt-\sqrt{\eta\nu}\,x_tz_t\,dW_t \\
	dy_t &= -\tfrac{1}{2}\nu y_t\,dt-\sqrt{\eta\nu}\,y_tz_t\,dW_t \\
	dz_t &= -u(t)x_t+\sqrt{\eta\nu}\,(1-z_t^2)\,dW_t.
\end{split}
\end{equation}
The idea is to first fix a feedback control law like the following affine function of the Bloch vector \cite{vanhandel-feedback}:
\beq u(t)=-\um(1+z_t)+2 x_t .\eeq
The choice stems from geometric considerations: the dependence by both $z_t$ and $x_t$ and the particular choice of parameters ensures that the control is zero at the target state, but not so on the antipodal state, i.e. the other eigenvector of $\sigma_z$ that would be stationary for the free dynamics.
Next, one can use existing numerical algorithms to find a polynomial Lyapunov function $V(\rho)$ such that $\LL V(\rho)$ can be decomposed as a (negative) sum of squares, where $\LL$ is the infinitesimal generator associated to the diffusion \eqref{SME5} \cite{kushner}. Given the previous analysis of the invariants for the uncontrolled diffusions and invoking the LaSalle-type results of \cite{kushner}, the effectiveness of the control law can be readily proven. 


There are no systematic and efficient ways of designing stabilizing controls for arbitrary SMEs. The problem has however been solved for a class of $N$-level spin systems, by resorting to a ``patched'' control law \cite{mirrahimi-stabilization}. Consider a system of the form \eqref{eq:SME}, where $H_1=F_y,$ and $L=F_z,$ where $F_y$ and $F_z$ are $N$-level spin operators \cite{mirrahimi-stabilization,sakurai}.

\begin{theorem}
Let the target state $\rho_{d}$ correspond to an eigenstate of $F_z$. Consider dynamics driven by \eqref{eq:SMEcont} and the following control law:
\begin{enumerate}
\item
$u_{t}=-\text{tr}(i[F_{y}, \rho_{t}]\rho_{d})$ if $\text{tr}(\rho_{t}\rho_{d})\geq
\gamma$.
\item $u_{t}=1$ if $\text{tr}(\rho_{t}\rho_{d})\leq
\gamma/2$.
\item If $\rho_{t} \in \mathcal{B}=\{\rho: \gamma/2 < \text{tr}(\rho\rho_{d})< \gamma\}$,
then $u_{t}=-\text{tr}(i[F_{y}, \rho_{t}]\rho_{d})$ if $\rho_{t}$ last entered $\mathcal{B}$ through the boundary
$\text{tr}(\rho\rho_{d})=\gamma$, and $u_{t}=1$ otherwise.
\end{enumerate}
Then there exists a $\gamma >0$ such that $u_{t}$ globally
stabilizes \eqref{eq:SMEcont} around $\rho_{d}$ and
$\mathbb{E}(\rho_{t})\rightarrow \rho_{d}$ as $t \rightarrow \infty$.
\end{theorem}

The idea underlying the result is to employ a stochastic Lyapunov-based feedback law in a neighborhood of the target state, and ``destabilize'' the rest of the state space via a constant control. For systems of the form \eqref{eq:SMEcont}, not just stabilization problems but a number of optimal feedback control problems have also been considered, see {\em e.g.} \cite{belavkin-report,mancini-wiseman,jacobs-viscosity}.

\subsection{Quantum optical networks}
\label{sec:lin-feedb-net}

\subsubsection{Basic concatenation rules and reducible networks}
One of the main advantages with quantum systems where the variables of interest are represented by traveling light fields is that, with good approximation, the interactions between the field and cavities, beam-splitters, mirrors, and any other component of the network are {\em local in space and time}. This allows one to define quantum {\em input and output} signals from the elements of the network, and thus interconnect a number of these in order to design {\em e.g.} some desired input-output behavior. A comprehensive review of this approach goes well beyond the scope of this tutorial introduction, so we limit ourselves to a sketch of the basic formalism for describing these networks.
Recall that the key equation for describing the field-system interaction {\em without measurements} is \eqref{QSDE}, which is completely specified by the (operator) parameters $\mathbf{G}=(S,L,H).$ This corresponds to a SISO block for our network. If we now want to couple multiple systems through input and output fields in a network, we can construct a joint model using two simple concatenation rules:

{\em Connection in series:}  If the input of a system associated to parameters $\mathbf{G}_1=(S_1,L_1,H_1)$ feeds its output field into a second system $\mathbf{G}_2=(S_2,L_2,H_2)$, the joint system can be equivalently described as a single system $\mathbf{G}=\mathbf{G}_2 \triangleleft \mathbf{G}_1,$ where
\[\mathbf{G}=(S_2S_1,\,L_2+S_2L_1,\,H_1+H_2+(L_2^\dag S_2L_1)^S),\]
$X^S$ indicating the skew-Hermitian part of $X.$

{\em Concatenation:} If a system associated to parameters $\mathbf{G}_1=(S_1,L_1,H_1)$ is combined with another system $\mathbf{G}_2=(S_2,L_2,H_2)$, the joint system can be described as a single system 
$\mathbf{G}=\mathbf{G}_1 \boxplus
\mathbf{G}_2$ where
\[\mathbf{G}=\left(
\left(
\begin{array}{cc}
S_1 &  0    \\
0   & S_2    
\end{array}
\right)
,\,\left(
\begin{array}{c}
L_1   \\
L_2    
\end{array}\right),H_1+H_2\right).\]
A network of systems that can be described only through series connections between  subsystems of a concatenation is called a {\em reducible network.}  Not all networks are reducible, however, and a general Hamiltonian theory has been presented in \cite{james-commmath}. An approach for formulating control design problems in this setting, based on a quantum version of the passivity approach of Willems, has been developed in \cite{james-passivity}.

%
%

\subsubsection{Linear networks}

Consider a quantum system described by a quantum harmonic oscillator, that is, with some position and momentum operators satisfying the canonical commutation relations and a quadratic Hamiltonian (see Appendix \ref{qosc} for a brief review of the system), and assume that the initial state is a coherent state.  If we now further assume $S$ and $L$ to be scalar, the state remains coherent throughout the evolution, and the input-output relation \eqref{inputoutput} becomes time-invariant: $dB_{out}=SdB_{in}+Ldt.$

The idea can be generalized to a network of these {\em linear} quantum systems, obtaining a model of the form:
\begin{equation}\label{LQSE}
\begin{array}{l}
dx(t)=Ax(t)dt+Bdw(t); \ x(0)=x_{0} \\ dy(t)=Cx(t)dt+Ddw(t),
\end{array}
\end{equation}
where  $x(t)=[x_{1}(t) \dots x_{n}(t)]^{T}$ is a
vector of self-adjoint noncommutative {\em operators} on an appropriate Hilbert space, $dw(t)$ is the (noncommutative) increment of the vector of the input signals, $dy(t)$ of the output signals,
and $A$, $B$, $C$ and $D$ are matrices of appropriate dimensions.

A standard form can be introduced, in which the initial condition $x(0)=x_{0}$ of the system is taken to satisfy the commutation relations
\begin{equation}
[x_{j}(0), x_{k}(0)]=2i \delta_{jk}J,\quad j,k=1, \dots, n,
\end{equation}
where $J=[\begin{smallmatrix}0 && 1
\\ -1 && 0\end{smallmatrix}]$. The input $w$ is further assumed to admit the
decomposition
\begin{equation}\label{noiseinput}
dw(t)=\beta_{w}(t)dt+d\tilde{w}(t)
\end{equation}
where $\tilde{w}(t)$ is the noise part of $w(t)$ and $\beta_{w}(t)$ is the control to be designed.
The noise $\tilde{w}(t)$ is a vector of
self-adjoint noncommutative processes whose increment satisfies the It$\hat{\mbox{o}}$-like multiplication rule
$
d\tilde{w}(t)d\tilde{w}^{T}(t)=F_{\tilde{w}}dt,
$ where $F_{\tilde{w}}$ is a positive semidefinite Hermitian matrix. 

Many results from optimal and robust linear control theory have been adapted to this class of models \cite{james-2008,gough-linear,nurdin-coherent}, one of the key difficulties being to ensure the {\em physical realizability} of the controller one obtains from the numerical optimization procedures. One of the most relevant features of the approach is arguably the possibility of systematically designing and realizing {\em quantum feedback controllers}, or coherent controllers \cite{nurdin-coherent}:  namely, one can design a new {\em quantum} network whose output is exactly the (operator-valued) control $\beta_{w}(t).$ A typical control problem is the minimization of a quadratic functional
\[ {\cal J}(z)=\int_0^{t_f} \langle z(t)^\dag Q^\dag Q z(t) \rangle dt, \]
where $\langle \cdot \rangle$ denotes the expectation, and the ``augmented state" $z(t)$ includes the internal state of both the system and the controller.
Despite the linearity of the interconnected system, the optimization procedures are complicated by the nature of the physical constraints, making the problem non-convex. 
Recent numerical results in \cite{nurdin-coherent} showed that a coherent controller can outperform a (linear) classical one in the quantum LQG problem, suggesting that fully-quantum controllers are not just an intriguing theoretical challenge, but a potential advantage in quantum control theory. 

\section{A brief outlook}\label{outlook}

Manipulating physical systems at the quantum scale poses great challenges to the control engineer. A huge research effort has been spent in the last decade in the attempt of developing the early approaches to quantum control into a sound mathematical theory, as well as in devising techniques tailored to the most diverse experimental settings.
However, a lot remains to be done, both connecting the existing pieces of theory into an organic framework and providing viable solutions to old and new open problems. We provide in the following a short list of some promising research directions and open problems of theoretical interest.

\noindent {\em Control of open quantum systems:} As we have seen in the study of the controllability of the MME, the effect of noise can be destructive, drastically limiting the capabilities of the most widely-used, open-loop control designs. In order to be able to manipulate a quantum system on a sufficiently long time scale, e.g. for implementing a complex quantum algorithm, the system must be endowed with mechanisms able to fight the action of the noise. Beside open-loop control method \cite{viola-gates}, more research is needed in order to assess the potential of feedback methods for error correction \cite{ahn-feedback} or for engineering protected codes \cite{ticozzi-markovian}.
We believe that the most interesting and compelling developments in the theory of quantum control will regard open systems, in particular the systematic analysis of synergistic interactions between the system and its environment, be it a properly engineered interaction or a measurement apparatus 
\cite{Lloyd2002,Wiseman11Squinting}. 

{\noindent \em Quantum controllers for quantum systems:}
The idea of using a quantum system as a ``controller'' for another quantum system, originally formulated in \cite{wiseman-alloptical,lloyd-coherent}, has been recently developed in the framework of quantum optical networks. In this case, it has been shown that the feedback quantum controllers can potentially have better performances than their classical counterparts.
Hence, new research is needed in order to assess the potential of quantum controllers, to provide a unified view on the problem, and to develop systematic design procedures.

{\noindent \em Feedback control of non-Markovian models:}
While the literature on open-loop control ideas based on dynamical decoupling is vast, feedback control of non-Markovian control models remain relatively unexplored (at least form a control-theoretic viewpoint). The master equation approach is receiving growing attention in the mathematical physics literature, especially in determining standard forms and quantifying the non-Markovianity of the evolution, and its potential application to ``quantum biology'' problems makes those models a natural candidate for control oriented studies. Recent developments on non-Markovian stochastic dynamics make those a natural object for further research on feedback control \cite{Barc10}.


\noindent{\em Large scale systems and quantum information processing:}
The application of quantum control techniques to high-dimensional systems, or large ``networks'' of simpler systems, offers a number of formidable challenges to the control engineer. While the improvement of numerical optimal control algorithms is certainly one of the needed steps in this direction, we are facing some more fundamental theoretical issues. For example, it is by now well-known that controllability is connected to the possibility of breaking all (dynamical) symmetries and that most multipartite quantum systems can be controlled by means of a few ``actuators'' \cite{Wang2011Symmetry,Zeier1}, but a theory of optimal placement of these actuators is still missing \cite{Burgarth07Full}. 
Another set of problems is posed to the quantum control community by quantum information processing techniques, most notably quantum computing.
In this field, the system is a register comprising a large number of (logical) qubits, and the need is for effective and {\em scalable} algorithms for entanglement generation, gate synthesis with high fidelity, and noise suppression. Our impression is that the most interesting developments will be in techniques that go beyond Hamiltonian control 
\cite{barreiro}: stringent locality constraints, and the necessity of incorporating more control techniques to tackle a single task, make the challenge formidable, and yet extremely intriguing.

%
%

\appendix

\subsection{Notations} \label{notations}

Throughout the tutorial we will make use of Dirac's notation, inherited from the physics literature. A quick review of this and other relevant concepts is provided here. Some standard references for quantum mechanics are {\em e.g.} \cite{sakurai,nielsen-chuang}. We will use $i$ for the imaginary unit, and $c^*$ will stand for the complex-conjugate of $c\in\C$.

{\em Spaces and Operators:} Consider a finite-dimensional separable Hilbert space $\Hi$ over the complex field $\C$, $\Hi\simeq\C^{N}$. When it is convenient to explicitly specify the space dimension, we shall denote by $\Hi^N$ an $N$-dimensional Hilbert space. We denote with $\Hi^\dag$ its dual space, i.e. the Hilbert space of continuous linear functionals on $\Hi.$ 
$\bound(\Hi)\simeq \C^{N\times N}$ is the set of linear operators on $\Hi$: The natural inner product in $\bound(\Hi)$ is the Hilbert-Schmidt product, namely $\langle X,Y \rangle=\tr{X^\dag Y},$ where $\tr{} $ is the usual trace functional (which is well-defined in a finite dimensional setting).  We denote with $X^\dag$ the adjoint of $X$. $\herm(\Hi)$ is the subset of  $\bound(\Hi)$ of Hermitian operators. 

Given two linear operators $X,Y$ on $\Hi$, we  indicate with $[X,\,Y]:=XY-YX$ and $\{X,\,Y\}:=XY+YX,$ respectively  the {\em commutator} and the {\em anti-commutator} of the two operators. $I_{\tt subscript}$ stands for the identity operator on the space $\Hi_{\tt subscript}$, 
while $I_d, \, d\in\mathbb{N}$ is used to indicate the identity operator on $\C^d$.

{\em Dirac notation and Matrix representation:} Given a Hilbert space $\Hi$, $\ket{\psi}$ is used to denote a vector (called a {\em ket}) in $\Hi$, $\bra{\psi}$ denotes its dual (a {\em bra}), and $\braket{\psi}{\varphi}$ denotes the inner product. When $A$ is a linear operator on $\Hi,$ $A\ket{\psi}$ denotes the image of $\ket{\psi}$ through $A,$ and $\bra{\psi}A:=(A\ket{\psi})^\dag.$ Thus we have an unambiguous meaning for the notation $\bra{\psi}A\ket{\varphi}=\bra{\psi}(A\ket{\varphi})=(A\ket{\psi})^\dag\ket{\varphi})$.  The ``external'' product notation $\ket{\psi}\bra{\varphi}$ stands for the linear operator on $\Hi$ defined by
$\ket{\psi}\bra{\varphi}(\ket{\xi})=\ket{\psi}\braket{\varphi}{\xi},$
for all $\ket{\xi}\in\Hi.$
Hence, if $|\braket{\psi}{\psi}|=1,$  $\ket{\psi}\bra{\psi}$ represents the orthogonal projector onto the one dimensional subspace generated by $\ket{\psi}$. 

In matrix representation, $\ket{\psi}\in\Hi\simeq\C^N$ are represented by column vectors, so $\bra{\phi}\in\Hi^\dag\simeq\C^N$ are row vectors.  $X\in\bound(\Hi)\simeq\C^{N\times N}$ are thus represented by (generic) $N\times N$ complex matrices, the adjoint $X^\dag$ becomes the transpose conjugate of $X$, and hence $H\in\herm(\Hi)$ are represented by Hermitian matrices.

{\em Tensor-products and partial trace:} If $\Hi_1,\Hi_2$ are two Hilbert spaces, the notation $\Hi_1\otimes\Hi_2$  refers to the closed linear span of the tensor-product vectors $\ket{\psi_1}\otimes\ket{\psi_2},$ where $\ket{\psi}\in\Hi_1,\ket{\psi_2}\in\Hi_2.$ In matrix representation, it corresponds to the Kronecker product. 
Given $X_1\in{\frak B}(\Hi_1),X_2\in{\frak B}(\Hi_2),$ $X_1\otimes X_2$ is the linear operator on $\Hi_1\otimes\Hi_2,$ such that $X_1\otimes X_2(\ket{\psi_1}\otimes\ket{\psi_2})=X_1(\ket{\psi_1})\otimes X_2(\ket{\psi_2}),$ where $\ket{\psi_1}\in\Hi_1,\ket{\psi_2}\in\Hi_2,$ and extended to the whole tensor space by linearity.
The {\em partial trace} of $X$ over $\Hi_2$, $\trace_{2}(X)$, is the only linear functional on ${\mathfrak B}(\Hi_{1}\otimes\Hi_{2})$ such that for every $X_1\in{\mathfrak B}(\Hi_{1}),$ the above property holds. It can be defined starting from factorizable operators as:
$ \trace_2(X_1\otimes X_2)=X_1\trace(X_2),$
and it is extended to all $X_{12}\in{\frak B}(\Hi_1\otimes\Hi_2)$ by linearity. See {\em e.g.} \cite{nielsen-chuang} for a detailed discussion. 

\subsection{Quantum harmonic oscillator}\label{qosc}

Let us recall the key elements of the quantum oscillator model with the ``ladder'' operator method, originally due to Dirac, referring to \cite{sakurai} for a more complete treatment. An harmonic oscillator of frequency $\omega$ in quantum theory is an infinite-dimensional system with discrete energy levels $E_n=(n+\frac{1}{2})\omega,\,n=0,1,2,\ldots ,$ with corresponding energy eigenvectors $\ket{\psi_n}.$  
Let us define the {\em annihilation} and {\em creation} operators, $a$ and $a^\dag$ in terms of their action on the energy eigenbasis:
\[a\ket{\psi_n}:=\sqrt{n}\ket{\psi_{n-1}},\, a^\dag\ket{\psi_n}:=\sqrt{n+1}\ket{\psi_{n+1}}.\]
The definition reveals the reason for their names: $a$ lowers the energy level, subtracting an energy ``particle'' or quantum, while $A^\dag$ raises the energy level, adding a quantum. They also satisfy the {\em canonical commutation relation} $[a,a^\dag]=I.$

In terms of these operators, one can then define the counting, or {\em number} operator, $N:=a^\dag a$ which is diagonal in the energy eigenvector basis (hence also called the number basis) and $ N\ket{\psi_n}=n\ket{\psi_{n}}$. It is the observable associated to the number of quanta in the oscillator. 
The Hamiltonian can then be written as $H=\omega(N+\um I).$
In the Heisenberg picture, we have that
\[\ddt a=i[H,a]=-i\omega a,\quad\ddt a^\dag=i[H,a^\dag]=i\omega a^\dag,\]
so that $a(t)=e^{-i\omega t}a(0),\;a^\dag(t)=e^{i\omega t}a^\dag(0).$ Thus, they are energy eigen-operators for the evolution at each time.

An equivalent description in terms of the canonical position ($x$) and momentum ($p$) operators is recovered by defining
\[x(t)=\sqrt{\frac{2}{m\omega}}\frac{a(t)+a^\dag(t)}{2},\quad p(t)=\sqrt{{2}{m\omega}}\frac{a(t)-a^\dag(t)}{2i},\]
($m$ is the mass of the particle) satisfying the canonical commutation relation $[x,p]=i,$ so that the system Hamiltonian acquires the familiar quadratic form
\[H=\frac{p(t)^2}{2m}+\um m\omega^2x(t)^2.\]

An particularly interesting class of states of the harmonic oscillator are the coherent states. They are the eigenstates of the annihilation operator $a$. In the number basis, they are parametrized by the complex eigenvalue $\alpha$ and defined as 
\[\ket{\alpha}=e^{-\frac{|\alpha |^2}{2}}\sum_{n=0}^\infty \frac{\alpha^n}{\sqrt{n!}}\ket{\psi_n}.\]
For these states, the expectation of $N$ is precisely $|\alpha|^2,$ as is its variance. Under the evolution driven by the harmonic oscillator Hamiltonian $H=\omega(N+\um I)$ a coherent state remains coherent with the same amplitude and an oscillating phase, that is
$\ket{\alpha(t)}:=e^{-iH}\ket{\alpha(0)}=\ket{\alpha e^{-i(\omega+\um) t}}.$

\bibliographystyle{IEEEtran} 

\bibliography{biblio-claudio-6}

\begin{thebibliography}{100}
\providecommand{\url}[1]{#1}
\csname url@rmstyle\endcsname
\providecommand{\newblock}{\relax}
\providecommand{\bibinfo}[2]{#2}
\providecommand\BIBentrySTDinterwordspacing{\spaceskip=0pt\relax}
\providecommand\BIBentryALTinterwordstretchfactor{4}
\providecommand\BIBentryALTinterwordspacing{\spaceskip=\fontdimen2\font plus
\BIBentryALTinterwordstretchfactor\fontdimen3\font minus
  \fontdimen4\font\relax}
\providecommand\BIBforeignlanguage[2]{{%
\expandafter\ifx\csname l@#1\endcsname\relax
\typeout{** WARNING: IEEEtran.bst: No hyphenation pattern has been}%
\typeout{** loaded for the language `#1'. Using the pattern for}%
\typeout{** the default language instead.}%
\else
\language=\csname l@#1\endcsname
\fi
#2}}

\bibitem{nielsen-chuang}
M.~A. Nielsen and I.~L. Chuang, \emph{Quantum Computation and
  Information}.\hskip 1em plus 0.5em minus 0.4em\relax Cambridge University
  Press, Cambridge, 2002.

\bibitem{bs12}
A.~Butkovskii and Y.~Samoilenko, ``Control of quantum systems,'' \emph{Automat.
  Rem. Control}, vol.~40, pp. 485--502, and 629--645, 1979.

\bibitem{Huang1983}
G.~M. Huang, T.~J. Tarn, and J.~W. Clark, ``On the controllability of
  quantum-mechanical systems,'' \emph{J. Math. Phys.}, vol.~24, no.~11, pp.
  2608--2618, 1983.

\bibitem{Belavkin1983}
V.~P. Belavkin, ``Theory of control of observable quantum systems,''
  \emph{Automatica and Remote Control}, vol.~44, no.~2, pp. 178--188, 1983.

\bibitem{Peirce1988}
A.~Peirce, M.~Dahleh, and H.~Rabitz, ``Optimal control of quantum mechanical
  systems: Existence, numerical approximations, and applications,'' \emph{Phys.
  Rev. A}, vol.~37, p. 4950, 1988.

\bibitem{Tannor2}
D.~J. Tannor and S.~A. Rice, ``Control of selectivity of chemical reaction via
  control of wave packet evolution,'' \emph{J. Chem. Phys.}, vol.~83, pp.
  5013--5018, 1985.

\bibitem{brumer92laser}
P.~Brumer and M.~Shapiro, ``Laser control of molecular processes,''
  \emph{Annual Review of Physical Chemistry}, vol.~43, no.~1, pp. 257--282,
  1992.

\bibitem{Brif2010Control}
C.~Brif, R.~Chakrabarti, and H.~Rabitz, ``Control of quantum phenomena: past,
  present and future,'' \emph{New Journal of Physics}, vol.~12, no.~7, p.
  075008, 2010.

\bibitem{Dahleh1}
M.~Dahleh, A.~Peirce, H.~Rabitz, and V.~Ramakrishna, ``Control of molecular
  motion,'' \emph{Proceedings of the IEEE}, vol.~84, pp. 7--15, 1996.

\bibitem{Rabitz1}
H.~Rabitz, R.~{de Vivie-Riedle}, M.~Motzkus, and K.~Kompa, ``Whither the future
  of controlling quantum phenomena?'' \emph{Science}, vol. 288, pp. 824--828,
  2000.

\bibitem{Ernst1}
R.~R. Ernst, G.~Bodenhausen, and A.~Wokaun, \emph{Principles of Magnetic
  Resonance in One and Two Dimensions}.\hskip 1em plus 0.5em minus 0.4em\relax
  Clarendon Press, Oxford, UK, 1987.

\bibitem{Mehring1}
M.~Mehring, \emph{High resolution NMR in solids}.\hskip 1em plus 0.5em minus
  0.4em\relax Springer-Verlag, 1976.

\bibitem{Cory1modif}
D.~G. {Cory {\em et al.}}, ``{NMR} quantum information processing: achievements
  and prospects,'' \emph{Prog. Phys.}, vol.~48, p. 875, 2000.

\bibitem{viola-seminalDD}
L.~Viola and S.~Lloyd, ``Dynamical suppression of decoherence in two-state
  quantum systems,'' \emph{Phys. Rev. A}, vol.~58, p. 2733, 1998.

\bibitem{Khaneja3}
N.~Khaneja, R.~Brockett, and S.~J. Glaser, ``Time optimal control in spin
  systems,'' \emph{Phys. Rev. A}, vol.~63, p. 032308, 2001.

\bibitem{Haeberlen1}
U.~Haeberlen and J.~S. Waugh, ``Coherent averaging effect in magnetic
  resonance,'' \emph{Phys. Rev.}, vol. 175, pp. 453--467, 1968.

\bibitem{dolinar}
S.~J. Dolinar, ``An optimum receiver for the binary coherent state quantum
  channel,'' \emph{Research Laboratory of Electronics, MIT, Quarterly Progress
  Report}, vol. 111, p. 115, 1973.

\bibitem{helstrom}
C.~W. Helstrom, \emph{Quantum Detection and Estimation Theory}.\hskip 1em plus
  0.5em minus 0.4em\relax Academic, New York, 1976.

\bibitem{wiseman-adaptive}
H.~M. Wiseman, ``Adaptive phase measurements of optical modes: Going beyond the
  marginal $q$ distribution,'' \emph{Phys. Rev. Lett.}, vol.~75, pp.
  4587--4590, 1995.

\bibitem{armen}
M.~A. Armen, J.~K. Au, J.~K. Stockton, A.~C. Doherty, and H.~Mabuchi,
  ``Adaptive homodyne measurement of optical phase,'' \emph{Phys. Rev. Lett.},
  vol.~89, p. 133602, 2002.

\bibitem{geremia}
R.~L. Cook, P.~J. Martin, and J.~M. Geremia, ``Optical coherent state
  discrimination using a closed-loop quantum measurement,'' \emph{Nature}, vol.
  446, pp. 774--777, 2007.

\bibitem{wj85}
J.G.Walker and E.Jakeman, ``Optical dead time effects and sub-poissonian
  photo-electron counting statistics,'' \emph{Proc. Soc. Photo-Opt. Instrum.
  Eng.}, vol. 492, p. 274, 1985.

\bibitem{hy86}
H.~A. Haus and Y.~Yamamoto, ``Theory of feedback-generated squeezed states,''
  \emph{Phys. Rev. A}, vol.~34, pp. 270--292, 1986.

\bibitem{yim86}
Y.~Yamamoto, N.~Imoto, and S.~Machida, ``Amplitude squeezing in a semiconductor
  laser using quantum nondemolition measurement and negative feedback,''
  \emph{Phys. Rev. A}, vol.~33, pp. 3243--3261, 1986.

\bibitem{ssh}
J.~H. Shapiro, G.~Saplakoglu, S.~T. Ho, P.~Kumar, B.~E.~A. Saleh, and M.~C.
  Teich, ``Theory of light detection in the presence of feedback,'' \emph{J.
  Opt. Soc. Am. B}, vol.~4, p. 1604, 1987.

\bibitem{qed1}
H.~Mabuchi and A.~C. Doherty, ``{Cavity Quantum Electrodynamics: Coherence in
  Context},'' \emph{Science}, vol. 298, no. 5597, pp. 1372--1377, 2002.

\bibitem{qed2}
D.~A. Steck, K.~Jacobs, H.~Mabuchi, T.~Bhattacharya, and S.~Habib, ``Quantum
  feedback control of atomic motion in an optical cavity,'' \emph{Phys. Rev.
  Lett.}, vol.~92, no.~22, p. 223004, Jun 2004.

\bibitem{orozco1}
W.~P. Smith, J.~E. Reiner, L.~A. Orozco, S.~Kuhr, and H.~M. Wiseman, ``Capture
  and release of a conditional state of a cavity \mbox{QED} system by quantum
  feedback,'' \emph{Phys. Rev. Lett.}, vol.~89, no.~13, p. 133601, 2002.

\bibitem{vanhandel-feedback}
R.~van Handel, J.~K. Stockton, and H.~Mabuchi, ``Feedback control of quantum
  state reduction,'' \emph{IEEE Trans. Aut. Contr.}, vol.~50, no.~6, pp.
  768--780, 2005.

\bibitem{wiseman-book}
H.~M. Wiseman and G.~J. Milburn, \emph{Quantum Measurement and Control}.\hskip
  1em plus 0.5em minus 0.4em\relax Cambridge University Press, 2009.

\bibitem{petta-science-mod}
J.~R. {Petta {\em et al.}}, ``Coherent manipulation of coupled electron spins
  in semiconductor quantum dots,'' \emph{Science}, vol. 309, pp. 2180--2184,
  2005.

\bibitem{qdots1}
D.~Press, T.~D. Ladd, and B.~Z. .~Y. Yamamoto, ``Complete quantum control of a
  single quantum dot spin using ultrafast optical pulses,'' \emph{Nature}, vol.
  456, no. 7219, pp. 218--221, 2008.

\bibitem{nanomechanical1}
S.~Mancini, D.~Vitali, and P.~Tombesi, ``Optomechanical cooling of a
  macroscopic oscillator by homodyne feedback,'' \emph{Phys. Rev. Lett.},
  vol.~80, no.~4, pp. 688--691, Jan 1998.

\bibitem{nanomechanical2}
A.~Hopkins, K.~Jacobs, S.~Habib, and K.~Schwab, ``Feedback cooling of a
  nanomechanical resonator,'' \emph{Phys. Rev. B}, vol.~68, no.~23, p. 235328,
  Dec 2003.

\bibitem{nanomechanical3}
R.~Ruskov, K.~Schwab, and A.~N. Korotkov, ``Squeezing of a nanomechanical
  resonator by quantum nondemolition measurement and feedback,'' \emph{Phys.
  Rev. B}, vol.~71, no.~23, p. 235407, Jun 2005.

\bibitem{ions1}
L.-M. Duan, J.~I. Cirac, and P.~Zoller, ``{Geometric Manipulation of Trapped
  Ions for Quantum Computation},'' \emph{Science}, vol. 292, no. 5522, pp.
  1695--1697, 2001.

\bibitem{ions2-mod}
P.~{Bushev {\em et al.}}, ``Feedback cooling of a single trapped ion,''
  \emph{Phys. Rev. Lett.}, vol.~96, no.~4, p. 043003, Feb 2006.

\bibitem{josephson1}
J.~F. Ralph, E.~J. Griffith, T.~D. Clark, and M.~J. Everitt, ``Guidance and
  control in a josephson charge qubit,'' \emph{Phys. Rev. B}, vol.~70, no.~21,
  p. 214521, Dec 2004.

\bibitem{josephson2}
S.~Montangero, T.~C., and R.~Fazio, ``Robust optimal quantum gates for
  josephson charge qubits,'' \emph{Phys. Rev. Lett.}, vol.~99, no.~17, p.
  170501, Oct 2007.

\bibitem{cold0}
S.~Chu, ``{Cold atoms and quantum control},'' \emph{Nature}, vol. 416, no.
  5630, pp. 206--210, 2002.

\bibitem{cold1}
J.~I. Cirac and P.~Zoller, ``{How to Manipulate Cold Atoms},'' \emph{Science},
  vol. 301, no. 5630, pp. 176--177, 2003.

\bibitem{Jacques09-mod}
V.~{Jacques {\em et al.}}, ``Dynamic polarization of single nuclear spins by
  optical pumping of nitrogen-vacancy color centers in diamond at room
  temperature,'' \emph{Phys. Rev. Lett.}, vol. 102, no.~5, p. 057403, 2009.

\bibitem{Jiang09-mod}
L.~{Jiang {\em et al.}}, ``Repetitive readout of a single electronic spin via
  quantum logic with nuclear spin ancillae,'' \emph{Science}, vol. 326, no.
  5950, pp. 267--272, 2009.

\bibitem{gillet-mod}
G.~C. {Gillett {\em et al.}}, ``Experimental feedback control of quantum
  systems using weak measurements,'' \emph{Phys. Rev. Lett.}, vol. 104, p.
  080503, 2010.

\bibitem{polzik}
E.~S. Polzik, B.~Julsgaarda, J.~Sherson, and J.~L. SËrensen, ``Entanglement and
  quantum teleportation with multi-atom ensembles,'' \emph{Philos. Trans. R.
  Soc. London, Ser. A}, vol. 361, 2003.

\bibitem{hammerer}
K.~Hammerer, A.~S. S\o{}rensen, and E.~S. Polzik, ``Quantum interface between
  light and atomic ensembles,'' \emph{Rev. Mod. Phys.}, vol.~82, pp.
  1041--1093, 2010.

\bibitem{Mabuchi2005}
H.~Mabuchi and N.~Khaneja, ``Principles and applications of control in quantum
  systems,'' \emph{International Journal of Robust and Nonlinear Control},
  vol.~15, pp. 647--667, 2005.

\bibitem{dalessandro-book}
D.~D'Alessandro, \emph{Introduction to Quantum Control and Dynamics}, ser.
  Applied Mathematics \& Nonlinear Science.\hskip 1em plus 0.5em minus
  0.4em\relax Chapman \& Hall/CRC, 2007.

\bibitem{vanhandel-invitation}
L.~Bouten, R.~van Handel, and M.~R. James, ``A discrete invitation to quantum
  filtering and feedback control,'' \emph{SIAM Review}, vol.~51, no.~2, pp.
  239--316, 2009.

\bibitem{Dong-Petersen-survey}
D.~Dong and I.~Petersen, ``Quantum control theory and applications: a survey,''
  \emph{IET Control Theory Appl.}, vol.~4, no.~12, p. 2651 † 2671, 2010.

\bibitem{sakurai}
J.~J. Sakurai, \emph{Modern Quantum Mechanics}.\hskip 1em plus 0.5em minus
  0.4em\relax Addison-Wesley, New York, 1994.

\bibitem{Peres2}
A.~Peres, \emph{Quantum theory : concepts and methods}.\hskip 1em plus 0.5em
  minus 0.4em\relax Kluwer, 1993.

\bibitem{isham-qm}
C.~Isham, \emph{Lectures on Quantum Theory: Mathematical and Structural
  Foundations}.\hskip 1em plus 0.5em minus 0.4em\relax Imperial College Press,
  London, 1995.

\bibitem{parthasarathy}
K.~R. Parthasarathy, \emph{An Introduction to Quantum Stochastic Calculus},
  ser. Monographs in Mathematics.\hskip 1em plus 0.5em minus 0.4em\relax
  Birkhauser, Basel, 1992, vol.~85.

\bibitem{maassen-qp}
H.~Maassen, ``Quantum probability applied to the damped harmonic oscillator,''
  in \emph{Quantum Probability Communications}, S.~Attal and J.~Lindsay,
  Eds.\hskip 1em plus 0.5em minus 0.4em\relax World Scientific, 2003, pp.
  23--58.

\bibitem{alicki-lendi}
R.~Alicki and K.~Lendi, \emph{Quantum Dynamical Semigroups and
  Applications}.\hskip 1em plus 0.5em minus 0.4em\relax Springer-Verlag,
  Berlin, 1987.

\bibitem{petruccione}
H.~P. Breuer and F.~Petruccione, \emph{The Theory of Open Quantum
  Systems}.\hskip 1em plus 0.5em minus 0.4em\relax Oxford University Press, UK,
  2006.

\bibitem{Brockett3}
R.~Brockett, ``Systems theory on group manifolds and coset spaces,'' \emph{SIAM
  Journal on Control}, vol.~10, pp. 265--284, 1972.

\bibitem{Jurdjevic1}
V.~Jurdjevic, \emph{Geometric Control Theory}.\hskip 1em plus 0.5em minus
  0.4em\relax Cambridge, UK: Cambridge University Press, 1996.

\bibitem{Bengtsson1}
I.~Bengtsson and K.~Zyczkowski, \emph{Geometry of quantum states}.\hskip 1em
  plus 0.5em minus 0.4em\relax Cambridge University Press, 2006.

\bibitem{Lendi1}
K.~Lendi, ``Dynamical invariants for time-evolution of open quantum systems in
  finite dimension,'' \emph{J. of Phys. A: Math. Gen}, vol.~27, pp. 609--630,
  1994.

\bibitem{Schirmer7}
S.~G. Schirmer, T.~Zhang, and J.~V. Leahy, ``Orbits of quantum states and
  geometry of {B}loch vectors for {N}-level systems,'' \emph{J. Phys. A},
  vol.~37, pp. 1389--1402, 2004.

\bibitem{wiseman-milburn}
H.~M. Wiseman and G.~J. Milburn, ``Quantum theory of optical feedback via
  homodyne detection,'' \emph{Phys. Rev. Lett.}, vol.~70, no.~5, pp. 548--551,
  1993.

\bibitem{carmicheal}
H.~Carmichael, \emph{An Open Systems Approach to Quantum Optics}, ser. L.N. in
  Physics.\hskip 1em plus 0.5em minus 0.4em\relax Springer, Berlin, 1993,
  vol.~18.

\bibitem{lindblad}
G.~Lindblad, ``On the generators of quantum dynamical semigroups,'' \emph{Comm.
  Math. Phys.}, vol.~48, no.~2, pp. 119--130, 1976.

\bibitem{gorini-k-s}
V.~Gorini, A.~Kossakowski, and E.~Sudarshan, ``Completely positive dynamical
  semigroups of n-level systems,'' \emph{J. Math. Phys.}, vol.~17, no.~5, pp.
  821--825, 1976.

\bibitem{ziedler-functional}
E.~Zeidler, \emph{Applied Functional Analysis: Applications to Mathematical
  Physics}, ser. Applied mathematical sciences.\hskip 1em plus 0.5em minus
  0.4em\relax Springer Verlag New York, 1999, vol. 108.

\bibitem{schirmer-markovian}
S.~G. Schirmer and X.~Wang, ``Stabilizing open quantum systems by markovian
  reservoir engineering,'' \emph{Phys. Rev. A}, vol.~81, no.~6, p. 062306,
  2010.

\bibitem{ticozzi-markovian}
F.~Ticozzi and L.~Viola, ``Analysis and synthesis of attractive quantum
  {M}arkovian dynamics,'' \emph{Automatica}, vol.~45, pp. 2002--2009, 2009.

\bibitem{ticozzi-QDS}
------, ``Quantum {M}arkovian subsystems: Invariance, attractivity and
  control,'' \emph{IEEE Trans. Aut. Contr.}, vol.~53, no.~9, pp. 2048--2063,
  2008.

\bibitem{baumgartner-2}
B.~Baumgartner and H.~Narnhofer, ``Analysis of quantum semigroups with
  {GKS}--lindblad generators: {II}. general,'' \emph{Journal of Physics A:
  Mathematical and Theoretical}, vol.~41, no.~39, p. 395303, 2008.

\bibitem{Cla-contr-open1}
C.~Altafini, ``Controllability properties for finite dimensional quantum
  {M}arkovian master equations,'' \emph{J. Math. Phys.}, vol.~44, no.~6, pp.
  2357--2372, 2003.

\bibitem{wiseman-feedback}
H.~M. Wiseman, ``Quantum theory of continuous feedback,'' \emph{Phys. Rev. A},
  vol.~49, no.~3, pp. 2133--2150, 1994.

\bibitem{transferfunction-1}
M.~Yanagisawa and H.~Kimura, ``Transfer function approach to quantum
  control-part {I}: Dynamics of quantum feedback systems,'' \emph{IEEE Trans.
  Aut. Contr}, vol.~48, no.~12, pp. 2107 -- 2120, 2003.

\bibitem{gardiner-qn}
C.~W. Gardiner and P.~Zoller, \emph{Quantum Noise: A Handbook of Markovian and
  Non-Markovian Quantum Stochastic Methods with Applications to Quantum
  Optics}, 3rd~ed.\hskip 1em plus 0.5em minus 0.4em\relax Springer-Verlag, N.
  Y., 2004.

\bibitem{Barc09}
A.~Barchielli and M.~Gregoratti, \emph{Quantum Trajectories and Measurements in
  Continuous Time: The Diffusive Case}, ser. Lecture Notes in Physics,
  782.\hskip 1em plus 0.5em minus 0.4em\relax Springer, Berlin Heidelberg,
  2009.

\bibitem{vanhandel-filtering}
L.~Bouten, R.~van Handel, and M.~R. James, ``An introduction to quantum
  filtering,'' \emph{SIAM J. Control Optim.}, vol.~46, no.~2, pp. 2199--2241,
  2007.

\bibitem{belavkin-filtering}
V.~P. Belavkin, ``Quantum stochastic calculus and quantum nonlinear
  filtering,'' \emph{Journal of Multivariate Analysis}, vol.~42, pp. 171--201,
  1992.

\bibitem{hudson-fermions}
D.~Applebaum and R.~Hudson, ``Fermion itoÕs formulas and stochastic
  evolutions,'' \emph{Commun. Math. Phys.}, no.~96, p. 473496, 1984.

\bibitem{mirrahimi-stabilization}
M.~Mirrahimi and R.~V. Handel, ``Stabilizing feedback controls for quantum
  systems,'' \emph{SIAM J. Control Optim.}, vol.~46, pp. 445--467, 2007.

\bibitem{kushner}
H.~J. Kushner, \emph{Stochastic Stability and Control}.\hskip 1em plus 0.5em
  minus 0.4em\relax Academic Press, New York, 1967.

\bibitem{adler}
T.~A.~B. S.~L.~Adler, D. C.~Brody and L.~P. Hughston, ``Martingale models for
  quantum state reduction,'' \emph{Journal of Physics A: Mathematical and
  General}, vol.~34, pp. 8795--8820, 2001.

\bibitem{Doherty2000}
A.~C. Doherty, S.~Habib, K.~Jacobs, H.~Mabuchi, and S.~M. Tan, ``Quantum
  feedback control and classical control theory,'' \emph{Phys. Rev. A},
  vol.~62, p. 012105, 2000.

\bibitem{JamesLN}
M.~James, ``Lecture notes for {PHYS4003B},'' Australian National University,
  2007.

\bibitem{Jurdjevic3}
V.~Jurdjevic and H.~Sussmann, ``Control systems on {L}ie groups,''
  \emph{Journal of Differential Equations}, vol.~12, pp. 313--319, 1972.

\bibitem{Boothby3}
W.~Boothby and E.~Wilson, ``Determination of the transitivity of bilinear
  systems,'' \emph{SIAM J. Control Optim.}, vol.~17, pp. 212--221, 1979.

\bibitem{Albertini1}
F.~Albertini and D.~D'Alessandro, ``Notions of controllability for bilinear
  multilevel quantum systems,'' \emph{IEEE Trans. Autom. Contr.}, vol.~48, pp.
  1399--1403, 2003.

\bibitem{kuranishi1}
M.~Kuranishi, ``On everywhere dense imbeddings of free group on a {L}ie
  group,'' \emph{Nagoya Mathematical Journal}, vol.~2, pp. 63--71, 1951.

\bibitem{Jurdjevic2}
V.~Jurdjevic and G.~Sallet, ``Controllability properties of affine systems,''
  \emph{SIAM J. Control and Optimization}, vol.~22, pp. 501--508, 1984.

\bibitem{Ramakrishna4}
H.~R. V.~Ramakrishna, ``Relation between quantum computing and quantum
  controllability,'' \emph{Phys. Rev. A}, vol.~54, pp. 1715--1716, 1996.

\bibitem{Lloyd1}
S.~Lloyd, ``Almost any quantum logic gate is universal,'' \emph{Physical Review
  Letters}, vol.~75, pp. 346--349, 1995.

\bibitem{Turinici1}
G.~Turinici and H.~Rabitz, ``Quantum wave function controllability,''
  \emph{Chem. Phys.}, vol. 267, pp. 1--9, 2001.

\bibitem{Cla-contr-root1}
C.~Altafini, ``Controllability of quantum mechanical systems by root space
  decomposition of su({N}),'' \emph{J. Math. Phys.}, vol.~43, no.~5, pp.
  2051--2062, 2002.

\bibitem{SilvaLeite2}
F.~{Silva Leite} and P.~Crouch, ``Controllability on classical {L}ie groups,''
  \emph{Mathematics of Control, Signals and Systems}, vol.~1, pp. 31--42, 1988.

\bibitem{Marsden2}
J.~Marsden and T.~Ratiu, \emph{Introduction to Mechanics and Symmetry},
  2nd~ed., ser. Texts in Applied Mathematics.\hskip 1em plus 0.5em minus
  0.4em\relax Springer-Verlag, 1999, vol.~17.

\bibitem{Kurniawan07Dynamics}
I.~Kurniawan, G.~Dirr, and U.~Helmke, ``The dynamics of open quantum systems:
  Accessibility results,'' in \emph{Proceedings of the GAMM Annual Meeting},
  Zurich, 2007, pp. 4\,130\,045--4\,130\,046.

\bibitem{Dirr09Lie}
G.~Dirr, U.~Helmke, I.~Kurniawan, and T.~Schulte-Herbr{\"u}ggen,
  ``{L}ie-semigroup structures for reachability and control of open quantum
  systems: {K}ossakowski-{L}indblad generators from {L}ie wedges to {M}arkovian
  channels,'' \emph{Reports on Math. Physics}, vol.~64, pp. 93--121, 2009.

\bibitem{Yuan10Characterization}
H.~Yuan, ``Characterization of majorization monotone quantum dynamics,''
  \emph{IEEE Trans. Aut. Contr.}, vol.~55, no.~4, pp. 955 --959, 2010.

\bibitem{Cla-contr-open4}
C.~Altafini, ``Coherent control of open quantum dynamical systems,''
  \emph{Phys. Rev. A}, vol.~70, p. 062321, 2004.

\bibitem{viola-engineering}
S.~Lloyd and L.~Viola, ``Engineering quantum dynamics,'' \emph{Phys. Rev. A},
  vol.~65, pp. 010\,101:1--4, 2001.

\bibitem{poyatos}
J.~F. Poyatos, J.~I. Cirac, and P.~Zoller, ``Quantum reservoir engineering with
  laser cooled trapped ions,'' \emph{Phys. Rev. Lett.}, vol.~77, no.~23, pp.
  4728--4731, 1996.

\bibitem{davidovich}
A.~R.~R. Carvalho, P.~Milman, R.~L. de~Matos~Filho, and L.~Davidovich,
  ``Decoherence, pointer engineering, and quantum state protection,''
  \emph{Phys. Rev. Lett.}, vol.~86, no.~22, pp. 4988--4991, 2001.

\bibitem{Verstraete2009}
F.~Verstraete, M.~M. Wolf, and J.~I. Cirac, ``Quantum computation and
  quantum-state engineering driven by dissipation,'' \emph{Nature Physics},
  vol.~5, pp. 633 -- 636, 2009.

\bibitem{ticozzi-generic}
F.~Ticozzi, S.~G. Schirmer, and X.~Wang, ``Stabilizing quantum states by
  constructive design of open quantum dynamics,'' \emph{IEEE Trans. Aut.
  Contr.}, vol.~55, no.~12, pp. 2901 --2905, 2010.

\bibitem{Zhu1998385}
W.~Zhu and H.~Rabitz, ``A rapid monotonically convergent iteration algorithm
  for quantum optimal control over the expectation value of a positive definite
  operator,'' \emph{Journal of Chemical Physics}, vol. 109, no.~2, pp.
  385--391, 1998.

\bibitem{Maday20038191}
Y.~Maday and G.~Turinici, ``New formulations of monotonically convergent
  quantum control algorithms,'' \emph{Journal of Chemical Physics}, vol. 118,
  no.~18, pp. 8191--8196, 2003.

\bibitem{Krotov83Iteration}
V.~F. Krotov and I.~N. Feldman, ``Iteration method of solving the problems of
  optimal control,'' \emph{Eng. Cybern.}, vol.~21, p. 123, 1983.

\bibitem{Khaneja05optimal}
N.~Khaneja, T.~Reiss, C.~Kehletb, T.~Schulte-Herbr{\"u}ggen, and S.~J. Glaser,
  ``Optimal control of coupled spin dynamics: design of {NMR} pulse sequences
  by gradient ascent algorithms,'' \emph{J. Magn. Res.}, vol. 172, pp.
  296--305, 2005.

\bibitem{Machnes10Comparing}
S.~Machnes, U.~Sander, S.~Glaser, P.~de~Fouquieres, A.~Gruslys, S.~Schirmer,
  and T.~Schulte-Herbr{\"u}ggen, ``Comparing, optimising and benchmarking
  quantum control algorithms in a unifying programming framework,''
  \emph{arXiv:quant-ph/1011.4874}, 2010.

\bibitem{Boscain1}
U.~Boscain, G.~Charlot, J.-P. Gauthier, S.~Gu{\'e}rin, and H.-R. Jauslin,
  ``Optimal control in laser-induced population transfer for two- and
  three-level quantum systems,'' \emph{J. Math. Phys.}, vol.~43, no.~5, pp.
  2107--2132, 2002.

\bibitem{Ferrante1}
A.~Ferrante, M.~Pavon, and G.~Raccanelli, ``Driving the propagator of a spin
  system: a feedback approach,'' in \emph{Proc. of the 41st Conference on
  Decision and Control}, Las Vegas, NV, December 2002, pp. 46--50.

\bibitem{Mirrahimi2}
M.~Mirrahimi, P.~Rouchon, and G.~Turinici, ``Lyapunov control of bilinear
  {S}chr{\"o}dinger equations,'' \emph{Automatica}, vol.~41, pp. 1987--1994,
  2005.

\bibitem{Cla-qu-ens-feeb1}
C.~Altafini, ``Feedback stabilization of isospectral control systems on complex
  flag manifolds: application to quantum ensembles,'' \emph{IEEE Trans. Aut.
  Contr.}, vol.~52, no.~11, pp. 2019--2028, 2007.

\bibitem{schirmer-wang-2010a}
X.~Wang and S.~Schirmer, ``Analysis of {L}yapunov method for control of quantum
  states,'' \emph{IEEE Trans. Autom. Contr.}, vol.~55, no.~10, pp. 2259 --2270,
  2010.

\bibitem{Bhat1}
S.~P. Bhat and D.~S. Bernstein, ``A topological obstruction to continuous
  global stabilization of rotational motion and the unwinding phenomenon,''
  \emph{Systems Control Lett.}, vol.~39, no.~1, pp. 63--70, 2000.

\bibitem{vitali-flying}
S.~Damodarakurup, M.~Lucamarini, G.~D. Giuseppe, D.~Vitali, and P.~Tombesi,
  ``Experimental inhibition of decoherence on flying qubits via ``bang-bang''
  control,'' \emph{Phys. Rev. Lett.}, vol. 103, no.~4, p. 040502, Jul 2009.

\bibitem{viola-gates}
K.~Khodjasteh and L.~Viola, ``Dynamically error-corrected gates for universal
  quantum computation,'' \emph{Phys. Rev. Lett.}, vol. 102, no.~8, p. 080501,
  Feb 2009.

\bibitem{viola-dd}
L.~Viola, E.~Knill, and S.~Lloyd, ``Dynamical decoupling of open quantum
  system,'' \emph{Phys. Rev. Lett.}, vol.~82, no.~12, pp. 2417--2421, 1999.

\bibitem{ticozzi-feedbackDD}
F.~Ticozzi and L.~Viola, ``Single-bit feedback and quantum dynamical
  decoupling,'' \emph{Phys. Rev. A}, vol.~74, no.~5, pp. 052\,328:1--11, 2006.

\bibitem{magnus}
W.~Magnus, ``On the exponential solution of differential equations for a linear
  operator,'' \emph{Comm. Pure Appl. Math.}, vol.~7, pp. 649--673, 1954.

\bibitem{expcont2-mod}
A.~{Assion {\em et al.}}, ``{Control of Chemical Reactions by
  Feedback-Optimized Phase-Shaped Femtosecond Laser Pulses},'' \emph{Science},
  vol. 282, no. 5390, pp. 919--922, 1998.

\bibitem{expcont3}
T.~Weinacht, J.~Ahn, and P.~H. Bucksbaum, ``{Controlling the shape of a quantum
  wavefunction},'' \emph{{Nature}}, vol. {397}, no. {6716}, pp. {233--235},
  {1999}.

\bibitem{landscapes2}
R.~Chakrabarti and H.~Rabitz, ``Quantum control landscapes,''
  \emph{International Reviews in Physical Chemistry}, vol.~26, pp.
  671--735(65), 2007.

\bibitem{wiseman-alloptical}
H.~M. Wiseman and G.~J. Milburn, ``All-optical versus electro-optical
  quantum-limited feedback,'' \emph{Phys. Rev. A}, vol.~49, pp. 4110--4125,
  1994.

\bibitem{transferfunction-2}
M.~Yanagisawa and H.~Kimura, ``Transfer function approach to quantum
  control-part {II}: Control concepts and applications,'' \emph{IEEE Trans.
  Aut. Contr}, vol.~48, no.~12, pp. 2121 -- 2132, 2003.

\bibitem{james-2008}
M.~James, H.~Nurdin, and I.~Petersen, ``${H}^{\infty }$ control of linear
  quantum stochastic systems,'' \emph{IEEE Trans. Aut. Contr}, vol.~53, no.~8,
  pp. 1787 --1803, 2008.

\bibitem{gough-product}
J.~Gough and M.~R. James, ``The series product and its application to quantum
  feedforward and feedback networks,'' \emph{IEEE Trans. Aut. Contr}, vol.~54,
  no.~11, pp. 2530 --2544, 2009.

\bibitem{james-passivity}
M.~James and J.~Gough, ``Quantum dissipative systems and feedback control
  design by interconnection,'' \emph{IEEE Transactions on Automatic Control},
  vol.~55, no.~8, pp. 1806 --1821, 2010.

\bibitem{nurdin-coherent}
H.~Nurdin, M.~James, and I.~Petersen, ``Coherent quantum {LQG} control,''
  \emph{Automatica}, vol.~45, pp. 1837--1846, 2009.

\bibitem{HK97}
D.~B. Horoshko and S.~Y. Kilin, ``Direct detection feedback for preserving
  quantum coherence in an open cavity,'' \emph{Phys. Rev. Lett.}, vol.~78, pp.
  840--842, 1997.

\bibitem{TMW02}
L.~K. Thomsen, S.~Mancini, and H.~M. Wiseman, ``Spin squeezing via quantum
  feedback,'' \emph{Phys. Rev. A}, vol.~65, p. 061801, 2002.

\bibitem{ahn-feedback}
C.~Ahn, H.~M. Wiseman, and G.~J. Milburn, ``Quantum error correction for
  continuously detected errors,'' \emph{Phys. Rev. A}, vol.~67, no.~5, pp.
  052\,310:1--11, 2003.

\bibitem{WT01}
H.~M. Wiseman and L.~K. Thomsen, ``Reducing the linewidth of an atom laser by
  feedback,'' \emph{Phys. Rev. Lett.}, vol.~86, pp. 1143--1147, 2001.

\bibitem{wang-wiseman}
J.~Wang and H.~M. Wiseman, ``Feedback-stabilization of an arbitrary pure state
  of a two-level atom,'' \emph{Phys. Rev. A}, vol.~64, no.~6, pp.
  063\,810:1--9, 2001.

\bibitem{vanhandel-phd}
R.~van Handel, ``Filtering, stability, and robustness,'' Ph.D. dissertation,
  California Institute of Technology, 2006.

\bibitem{vanhandel-separation}
R.~van Handel and L.~Bouten, ``On the separation principle of quantum
  control,'' in \emph{Quantum Stochastics and Information: Statistics,
  Filtering and Control}, V.~P. Belavkin and M.~Guta, Eds.\hskip 1em plus 0.5em
  minus 0.4em\relax World Scientific, 2008.

\bibitem{Florchinger1994}
P.~Florchinger, ``A stochastic version of {J}urdjevic-{Q}uinn theorem,''
  \emph{Stochastic Anal. Appl.}, vol.~12, pp. 473--480, 1994.

\bibitem{Florchinger1995}
------, ``Lyapunov-like techniques for stochastic stability,'' \emph{SIAM J.
  Control and Optimization}, vol.~33, pp. 1151--1169, 1995.

\bibitem{belavkin-report}
V.~P. Belavkin, ``Measurement, filtering and control in quantum open dynamical
  systems,'' \emph{Rep. Math. Phys.}, vol.~43, pp. 405--425, 1999.

\bibitem{mancini-wiseman}
S.~Mancini and H.~M. Wiseman, ``Optimal control of entanglement via quantum
  feedback,'' \emph{Phys. Rev. A}, vol.~75, 2007.

\bibitem{jacobs-viscosity}
K.~Jacobs and A.~Shabani, ``Quantum feedback control: How to use veriÞcation
  theorems and viscosity solutions to find optimal protocols,'' \emph{Contemp.
  Phys.}, vol.~49, pp. 435--448, 2008.

\bibitem{james-commmath}
J.~Gough and M.~R. James, ``Quantum feedback networks: Hamiltonian
  formulation,'' \emph{Comm. Math. Phys.}, no. 287, pp. 1109--1132, 2009.

\bibitem{gough-linear}
J.~E. Gough, R.~Gohm, and M.~Yanagisawa, ``Linear quantum feedback networks,''
  \emph{Phys. Rev. A}, vol.~78, no.~6, p. 062104, 2008.

\bibitem{Lloyd2002}
S.~Lloyd and L.~Viola, ``Engineering quantum dynamics,'' \emph{Phys. Rev. A},
  vol.~65, p. 010101, 2002.

\bibitem{Wiseman11Squinting}
H.~M. Wiseman, ``Quantum control: Squinting at quantum systems,''
  \emph{Nature}, vol. 470, pp. 178--179, 2011.

\bibitem{lloyd-coherent}
S.~Lloyd, ``Coherent quantum feedback,'' \emph{Phys. Rev. A}, vol.~62, p.
  022108, 2000.

\bibitem{Barc10}
C.~P. A.~Barchielli, P. Di~Tella and F.~Petruccione, ``Stochastic schr\"odinger
  equations and memory,'' \emph{arXiv:1006.3647}, 2010.

\bibitem{Wang2011Symmetry}
X.~Wang, P.~Pemberton-Ross, and S.~G. Schirmer, ``Symmetry and subspace
  controllability for spin networks with a single-node control,'' \emph{IEEE
  Tr. Autom. Contr.}, p. to appear, 2011.

\bibitem{Zeier1}
R.~Zeier and T.~Schulte-Herbrueggen, ``Symmetry principles in quantum systems
  theory,'' \emph{Preprint quant-ph/1012.5256}, 2011.

\bibitem{Burgarth07Full}
D.~Burgarth and V.~Giovannetti, ``Full control by locally induced relaxation,''
  \emph{Phys. Rev. Lett.}, vol.~99, p. 100501, 2007.

\bibitem{barreiro}
J.~T. {Barreiro {\em et al.}}, ``An open-system quantum simulator with trapped
  ions,'' \emph{Nature}, vol. 470, pp. 486--491, 2011.

\end{thebibliography}
%
%
%
%
%
%

\end{document}